\documentclass[11pt]{article}
\usepackage{a4wide,amsfonts,theorem,float,verbatim}
{
\theoremstyle{break}
\theorembodyfont{\normalfont}
\newtheorem{theorem}{Theorem}[section]
\newtheorem{example}{Example}[section]
\newtheorem{corollary}{Corollary}[section]
\newtheorem{lemma}{Lemma}[section]
\newtheorem{definition}{Definition}[section]
\newtheorem{proposition}{Proposition}[section]
}
\newenvironment{proof}{\list{}{\setlength{\leftmargin}{0pt}}\item\relax
\noindent\emph{Proof.~}}{\endlist}

\newenvironment{remark}{\list{}{\setlength{\leftmargin}{0pt}}\item\relax
\noindent\emph{Remark.~}}{\endlist}

\newenvironment{remarks}{\list{}{\setlength{\leftmargin}{0pt}}\item\relax
\noindent\emph{Remarks.~}}{\endlist}

\makeatletter
\@addtoreset{equation}{section}
\@addtoreset{figure}{section}
\@addtoreset{table}{section}
\renewcommand\tableofcontents{%
\par\vspace*{0.3cm}
\noindent\emph{\sc\contentsname}
\par\vspace*{0.2cm}
    \@starttoc{toc}%
    }
\renewcommand\@dotsep{5}
\renewcommand\l@section{\@dottedtocline{1}{1.5em}{1.5em}}

\newdimen\z@ \z@=0pt
\def\m@th{\mathsurround=\z@}
\def\ialign{\everycr{}\tabskip\z@skip\halign}
\def\eqalign#1{\null\,\vcenter{\openup\jot\m@th
  \ialign{\strut\hfil$\displaystyle{##}$&$\displaystyle{{}##}$\hfil
      \crcr#1\crcr}}\,}
\makeatother
\newcommand\g{\ensuremath{\mathcal{G}}}
\newcommand\R{\ensuremath{\mathbb{R}}}
\def\ie{\emph{i.e.}}
\def\eg{\emph{e.g.}}

\begin{document}
\renewcommand{\thefootnote}{\fnsymbol{footnote}}

\begin{flushright}
\vspace*{-2cm}
FTUV/98-4\qquad IFIC/98-4
\\
January, 1998
\\[1cm]
\end{flushright}
\begin{center}
\begin{Large}
\bfseries{An introduction to some novel applications of Lie algebra cohomology
in mathematics and physics\footnote{To appear in the Proceedings of the VI Fall
Workshop on \emph{Geometry and physics} (Salamanca, September 1997).}}
\\[0.5cm]
\end{Large}
\begin{large}
J.~A. de Azc\'arraga$\dagger$,
J.~M.~Izquierdo$\ddagger$
and J.~C.~P\'erez Bueno$\dagger$
\end{large}
\\[0.3cm]
\begin{it}
$\dagger$ Departamento de F\'{\i}sica Te\'orica, Univ. de Valencia
\\
and IFIC, Centro Mixto Univ. de Valencia-CSIC,
\\
E--46100 Burjassot (Valencia), Spain.
\\[0.2cm]
$\ddagger$ Departamento de F\'{\i}sica Te\'orica, Universidad de Valladolid\\
E--47011, Valladolid, Spain
\end{it}
\end{center}

\begin{abstract}
After a self-contained introduction to Lie algebra cohomology, we present some
recent applications in mathematics and in physics.
\tableofcontents
\end{abstract}

\setcounter{footnote}{0}
\renewcommand{\thefootnote}{\arabic{footnote}}

\section{Preliminaries: $L_X,\ i_X,\ d$}
\label{sec2}
Let us briefly recall here some basic definitions and formulae which will be
useful later.
Consider a uniparametric group of diffeomorphisms of a manifold $M$,
$e^{X}:M\rightarrow M$, which takes a point
$x\in M$  of local coordinates $\{ x^i\}$ to
$x{'}^i\simeq x^i+\epsilon^i(x)$ ($= x^i+X^i(x)$).
Scalars and (covariant, say) tensors $t_q$ ($q=0,1,2,\dots$)
transform as follows
\begin{equation}
\phi{'}(x{'})=\phi(x)\quad,\quad
t{'}_i(x{'})=t_j(x)\frac{\partial x^j}{\partial x{'}^i}
\quad,\quad
t{'}_{i_1i_2}(x{'})=t_{j_1j_2}(x)\frac{\partial x^{j_1}}{\partial x{'}^{i_1}}
\frac{\partial x^{j_2}}{\partial x{'}^{i_2}}
\quad
\dots
\quad.
\label{prelc}
\end{equation}
In physics it is customary to define `local' variations, which compare the
transformed and original tensors at the same point $x$:
\begin{equation}
    \delta\phi(x)\equiv \phi{'}(x)-\phi(x)\quad , \quad
    \delta t_i(x)\equiv t{'}_i(x)-t_i(x)\quad ,\quad \dots \quad.
\label{preld}
\end{equation}
Then, the first order variation defines the Lie derivative:
\begin{equation}
    \eqalign{
\delta_\epsilon\psi=-\epsilon^j(x)\partial_j\psi(x):=-L_X\psi\quad ,\quad
(\delta_\epsilon t)_i=-(\epsilon^j\partial_j t_i+(\partial_i\epsilon^j)t_j)
:=-(L_X t)_i\quad ,
\cr
(\delta_\epsilon t)_{i_1i_2}=-(\epsilon^j\partial_j t_{i_1i_2}+
(\partial_{i_1}\epsilon^j)t_{ji_2}+(\partial_{i_2}\epsilon^j)t_{i_1j}):=
-(L_X t)_{i_1i_2}\quad .\cr}
\label{prele}
\end{equation}
Eqs. (\ref{prele}) motivate the following general definition:

\begin{definition}[{\normalfont\textit{Lie derivative}}{}]
\label{def2.1}
Let $\alpha$ be a (covariant, say) $q$ tensor on $M$,
$\alpha(x)=\alpha_{i_1\dots i_q}dx^{i_1}\otimes\dots\otimes dx^{i_q}$,
and $X=X^k\frac{\partial}{\partial x^k}$ a vector field $X\in\mathfrak{X}(M)$.
The \emph{Lie derivative} $L_X$
of $\alpha$ with respect to
$X$ is locally given by
\begin{equation}
(L_X\alpha)_{i_1\dots i_q}=X^k
\frac{\partial \alpha_{i_1\dots i_q}}{\partial x^k}+
\alpha_{ki_2\dots i_q}\frac{\partial X^k}{\partial x^{i_1}}+\dots +
    \alpha_{i_1\dots i_{q-1}k}\frac{\partial X^k}{\partial x^{i_q}}\quad
.\label{prelf}
\end{equation}
On a $q$-form $\displaystyle\alpha(x)=\frac{1}{q!}
\alpha_{i_1\dots i_q}dx^{i_1}\wedge\dots\wedge dx^{i_q}$,
$\alpha\in\wedge_q(M)$, $L_Y\alpha$ is defined by
\begin{equation}
   (L_Y\alpha)(X_{i_1},\dots,X_{i_q}):=Y\cdot\alpha(X_{i_1},\dots,X_{i_q})-
   \sum^q_{i=1}\alpha(X_{i_1},\dots,[Y,X_i],\dots, X_{i_q})
       \quad ;                                              \label{prelg}
\end{equation}
on vector fields, $L_X Y = [X,Y]$.
The action of $L_X$ on tensors of any type $t_q^p$ may be found using that
$L_X$ is a derivation,
\begin{equation}
    L_X(t\otimes t{'})=(L_Xt)\otimes t{'}+t\otimes L_X t{'}  \label{prelh}
\quad.
\end{equation}
\end{definition}

\begin{definition}[{\normalfont\textit{Exterior derivative}}{}]
The \emph{exterior derivative} $d$ is a derivation of degree +1,
$d:\wedge_q(M) \to \wedge_{q+1}(M)$;
it satisfies Leibniz's rule,
\begin{equation}
    d(\alpha\wedge\beta)=d\alpha\wedge\beta+(-1)^q\alpha\wedge d\beta
\quad,\quad
\alpha\in\Lambda_q\quad,
\label{preli}
\end{equation}
and is nilpotent, $d^2=0$. On the $q$-form above,
it is locally defined by
\begin{equation}
   d\alpha=\frac{1}{q!}\frac{\partial\alpha_{i_1\dots i_q}}{\partial
x^j}dx^j\wedge
    dx^{i_1}\wedge\dots\wedge dx^{i_q}\quad .              \label{prelj}
\end{equation}
The coordinate-free expression for the action of $d$ is (Palais formula)
\begin{equation}
    \eqalign{
     (d\alpha)(X_1,&\dots, X_q,X_{q+1}):=\sum^{q+1}_{i=1}(-1)^{i+1}
      X_i\cdot \alpha(X_1,\dots,{\hat X}_i,\dots,X_{q+1})\cr
     &+\sum_{i<j}(-1)^{i+j}\alpha([X_i,X_j],X_1,\dots,{\hat X}_i,\dots,
     {\hat X}_j,\dots,X_{q+1})\quad .\cr}                  \label{prelk}
\end{equation}
In particular, when $\alpha$ is a one-form,
\begin{equation}
   d\alpha(X_1,X_2)=X_1\cdot\alpha(X_2)-X_2\cdot\alpha(X_1)-
     \alpha([X_1,X_2])\quad .                              \label{prell}
\end{equation}
\end{definition}

\begin{definition}[{\normalfont\textit{Inner product}}{}]
The \emph{inner product} $i_X$ is the derivation of degree $-1$
defined by
\begin{equation}
(i_X\alpha)(X_1,\dots,X_{q-1})=\alpha(X,X_1,\dots,X_{q-1})\quad .
\label{prelm}
\end{equation}
On forms (Cartan decomposition of $L_X$),
\begin{equation}
     L_X=i_Xd+di_X\quad ,                                      \label{preln}
\end{equation}
from which $[L_X,d]=0$ follows trivially.
Other useful identity is
\begin{equation}
[L_X,i_Y]= i_{[X,Y]} \quad; \label{prelo}
\end{equation}
from (\ref{preln}) and (\ref{prelo}) it is easy to deduce that
$[L_X, L_Y] = L_{[X,Y]}$.
\end{definition}

\section{Elementary differential geometry on Lie groups}
\label{sec3}

Let $G$ be a Lie group and let $L_{g'} g =g'g = R_g g'$ ($g',\ g\in G$) be the
left and right actions $G\times G\to G$ with obvious notation.
The left (right) invariant vector fields LIVF (RIVF) on $G$ reproduce the
commutator of the Lie algebra \g\ of $G$
\begin{equation}
[X_{(i)}^L (g), X_{(j)}^L (g)] = C_{ij}^k X_{(k)}^L (g)
\quad,\quad
[X_{(i)}^R (g), X_{(j)}^R (g)] = - C_{ij}^k X_{(k)}^R (g)
\quad,\quad
C^\rho_{[i_1 i_2} C_{\rho i_3]}^\sigma = 0\quad,
\label{3jacobi}
\end{equation}
where the square bracket $[\ ]$ in the Jacobi identity (JI) means
antisymmetrization of the
indices $i_1, i_2, i_3$.
In terms of the Lie derivative,
the $L$- ($R$-) invariance conditions read\footnote{The
superindex L (R) in the fields
refers to the left (right) invariance of them;
LIVF (RIVF) generate right (left)
translations.}
\begin{equation}
L_{X_{(j)}^R (g)} X_{(i)}^L (g) = [X_{(j)}^R (g), X_{(i)}^L (g)] =0
\quad,
\quad
L_{X_{(i)}^L (g)} X_{(j)}^R (g) = [X_{(i)}^L (g), X_{(j)}^R (g)] =0
\quad.
\end{equation}

Let $\omega^{L(i)}(g)\in \wedge_1(G)$ be
the basis of LI one-forms
dual to a basis of \g\ given by LIVF
($\omega^{L(i)}(g)(X_{L(j)}(g)) =
\delta^i_j$).
Using (\ref{prell}), we get the Maurer-Cartan (MC) equations
\begin{equation}
d \omega^{L\,(i)}(g) = - {1\over 2} C^i_{jk} \omega^{L\,(j)}(g) \wedge
\omega^{L\,(k)}(g) \quad.
\label{3MC}
\end{equation}
In the language of forms, the JI in (\ref{3jacobi}) follows from $d^2=0$.
If the $q$-form $\alpha$ is LI
\begin{equation}
d\alpha^L (X^L_{i_1},\dots,X^L_{i_{q+1}}) =
\sum_{s<t}(-1)^{s+t}\alpha^L([X^L_{i_s},X^L_{i_t}],
X^L_{i_1},\dots,{\hat X}^L_{i_s},\dots,{\hat X}^L_{i_t},\dots,X^L_{i_{q+1}})
\quad,
\label{3diff}
\end{equation}
since $\alpha^L (X^L_1,\dots,{\hat X}^L_i,\dots,X^L_{q+1})$ in (\ref{prelk})
is constant and does not contribute\footnote{
{}From now on we shall assume that
vector fields and forms are left invariant (\ie, $X\in\mathfrak{X}^L(G)$, etc.)
and drop the superindex $L$. Superindices $L,\ R$ will be used
to avoid confusion when both LI and RI vector fields appear.}.
To facilitate the comparison with the generalized $\widetilde d_{m}$ to be
introduced in Sec.~\ref{sec8}, we note here that, with
$\widetilde d_2\equiv -d$, eq. (\ref{3diff}) is equivalent to
\begin{equation}
\widetilde d_2 \alpha^L(X^L_{i_1},\dots,X^L_{i_{q+1}}) =
{1\over (2\cdot 2 -2)!}{1\over (q-1)!}
\varepsilon_{i_1\dots i_{q+1}}^{j_1\dots j_{q+1}}
\alpha^L([X^L_{j_1},X^L_{j_2}],X^L_{j_3},\dots,X^L_{i_{q+1}})\quad.
\label{3diffa}
\end{equation}

The MC equations may be written in a more compact way by introducing the
(canonical) $\g$-valued LI one-form $\theta$ on $G$,
$\theta(g) = \omega^{(i)}(g) X_{(i)}(g)$;
then, MC equations read
\begin{equation}
d\theta = -\theta \wedge \theta = - {1\over 2}  [\theta,\theta]
\label{3canonMC}
\end{equation}
since, for $\g$-valued forms,
$[\alpha,\beta] := \alpha^{(i)} \wedge \beta ^{(j)} \otimes [X_{(i)},X_{(j)}]$.

The transformation properties of $\omega^{(i)}(g)$ follow from (\ref{prelg}):
\begin{equation}
L_{X_{(i)}(g)} \omega^{(j)}(g) = -C_{ik}^j \omega^{(k)}(g)
\quad.
\label{3Lomega}
\end{equation}
For a general LI $q$-form
$\displaystyle
\alpha(g)={1\over q!}\alpha_{i_1\dots i_q}
\omega^{(i_1)}(g)\wedge\dots\wedge\omega^{(i_q)}(g)$
on $G$
\begin{equation}
L_{X_{(i)}(g)} \alpha(g) =
-\sum_{s=1}^q {1\over q!} C^{i_{s}}_{i k} \alpha_{i_1 \dots i_q}
\omega^{(i_1)}(g)\wedge\dots\wedge \widehat{\omega^{(i_s)}(g)}
\wedge \omega^{(k)}(g) \wedge\dots\wedge \omega^{(i_q)}(g) \quad.
\label{3Lalpha}
\end{equation}

\section{Lie algebra cohomology: a brief introduction}
\label{sec4}

\subsection{Lie algebra cohomology}
\label{sec4.1}

\begin{definition}
[{\normalfont\textit{$V$-valued $n$-dimensional cochains on $\cal G$}}{}]
\label{def4.1}
Let $\g$ be a Lie algebra and $V$ a vector space.
A $V$-valued $n$-cochain $\Omega_n $ on ${\cal G}$
is a skew-symmetric $n$-linear mapping
\begin{equation}
\Omega_n: \g \wedge \mathop{\cdots}\limits^n \wedge \g \to V \quad,
\quad
\Omega_n^A = {1 \over n!} \Omega_{i_1 \dots i_n}^A \omega^{i_1} \wedge \dots
\wedge \omega^{i_n}
\quad,
\label{4cochain}
\end{equation}
where $\{\omega^{(i)}\}$ is a basis of $\g^*$ and the superindex $A$ labels
the components in $V$.
The (abelian) group of all $n$-cochains is denoted by
$C^{n} ({\cal G}, V )$.
\end{definition}

\begin{definition}
[{\normalfont\textit{Coboundary operator} (for the left action $\rho$ of
$\g$ on $V$)}{}]
\label{def4.2}
Let $V$ be a left $\rho({\cal G})$-module,
where $\rho$ is a representation of the Lie algebra $\g$,
$\rho(X_i)^A_{.C} \rho (X_j)^C_{.B} - \rho(X_j)^A_{.C} \rho (X_i)^C_{.B}
= \rho([X_i,X_j])^A_{.B}$.
The coboundary operator
$s: C^{n} ({\cal G} , V) \to C^{n{+}1} ({\cal G}, V )$
is defined by
\begin{equation}
\eqalign{( s \Omega_n )^A \, ( X_{1} ,..., X_{n{+}1} ) & :=
\sum_{i{=}1}^{n{+}1}
(-)^{i + 1} \rho ( X_i )^A_{.B} \,( \Omega_n^B ( X_{1} ,..., {\hat X}_{i},...,
X_{n{+}1} ) )\cr
 &+ \sum_{{j,k=1 \atop j < k}}^{n{+}1} (-)^{j{+}k} \Omega_n^A ( [
X_{j} , X_{k} ] , \ X_{1} ,..., {\hat X}_{j} ,...,{\hat X}_{k},..., X_{n{+}1} )
\quad.\cr}
\label{4cobound}
\end{equation}
\end{definition}

\begin{proposition}
\label{prop4.1}
The \emph{Lie algebra cohomology} operator $s$ is nilpotent,
$s^2=0$.
\end{proposition}

\begin{proof}
Looking at (\ref{prelk}),
$s$ in (\ref{4cobound}) may be at this stage formally written as
\begin{equation}
(s)_{.B}^A = \delta_B ^A d + \rho(X_i)_{.B}^A \omega^i
\quad, \quad
(s= d+ \rho(X_i) \omega^i)
\quad.
\end{equation}
Then, the proposition follows from the fact that
\begin{equation}
\begin{array}{rl}
s^2 = & (\rho(X_i) \omega^i + d) (\rho(X_j) \omega^j + d)
= \rho(X_i) \rho(X_j) \omega^i \wedge \omega^j + \rho(X_i) \omega^i d +
\rho(X_j) d \omega^j + d^2
\\[0.25cm]
= & \displaystyle
- {1\over 2} \rho(X_j) C^j_{l k} \omega^l \wedge \omega^k
+ {1\over 2} [\rho(X_i), \rho(X_j) ]  \omega^i \wedge \omega^j = 0
\quad.
\end{array}
\end{equation}
\end{proof}

\begin{definition}[{\normalfont\textit{$n$-th cohomology group}}{}]
\label{def4.3}
An $n$-cochain $\Omega_n$ is a cocycle, $\Omega_n \in Z^n_\rho(\g,V)$, when
$s \Omega_n =0$.
If a cocycle $\Omega_n$ may be written as
$\Omega_n = s \Omega'_{n-1}$
in terms of an $(n-1)$-cochain $\Omega'_{n-1}$,
$\Omega_n$ is a coboundary, $\Omega_n \in B^n_\rho(\g,V)$.
The $n$-th Lie algebra cohomology group $H^n_\rho(\g,V)$ is defined by
\begin{equation}
H^n_\rho(\g,V) = Z^n_\rho(\g,V) / B^n_\rho(\g,V)\quad.
\end{equation}
\end{definition}

\subsection{Chevalley-Eilenberg formulation}
\label{sec4.2}

Let $V$ be $\R$, $\rho$ trivial. Then the first term in
(\ref{4cobound}) is not present and, on LI one-forms,
$s$ and $d$ act in the same manner.
Since there is a one-to-one correspondence between $n$-antisymmetric maps on
\g\
and LI $n$-forms on $G$, an $n$-cochain in $C^n(\g,\R)$ may also be given by
the LI form on $G$
\begin{equation}
\Omega(g)= {1\over n!} \Omega_{i_1\dots i_n}
\omega^{(i_1)}(g) \wedge \dots \wedge \omega^{(i_n)}(g)
\label{4CEco}
\end{equation}
and the Lie algebra cohomology coboundary operator is now $d$
\cite{Che.Eil:48} (the explicit dependence of the forms $\Omega(g)$,
$\omega^i(g)$ on $g$ will be omitted henceforth).

\begin{remark}
It should be noticed that the Lie algebra (CE) cohomology is in general
different from the de Rham
cohomology: a form $\beta$ on $G$ may be de Rham exact, $\beta=d\alpha$, but
the potential form $\alpha$
might not be a cochain \ie, a LI form\footnote{
This is, \eg, the case for certain forms which appear in the theory of
supersymmetric extended objects (superstrings).
This is not surprising due to the absence of global considerations in the
fermionic sector of supersymmetry.
The Lie algebra cohomology notions are easily extended to the `super Lie' case
(see \eg, \cite{Azc.Izq:95} for references on these subjects).}.
Nevertheless, for $G$ compact (see Proposition~\ref{prop7.1})
$H_{DR}(G) = H_0 (\g,\R)$.
\end{remark}

\begin{example}
Let $\g$ be the abelian two-dimensional algebra. The corresponding Lie group
is $\R^2$, which is de Rham trivial.
However, the translation algebra $\R^2$ has non-trivial Lie algebra cohomology,
and in fact it admits a non-trivial two-cocycle giving rise to
the three-dimensional Heisenberg-Weyl algebra.
\end{example}

\subsection{Whitehead's lemma for vector valued cohomology}

\begin{lemma}[{\normalfont\textit{Whitehead's lemma}}{}]
\label{lem4.1}
Let ${\cal G}$ be a finite-dimensional semisimple
Lie algebra over a field of characteristic zero and let $V$ be a
finite-dimensional irreducible $\rho ( {\cal G} )$-module such that $\rho (
{\cal G} ) V \not = 0$ ($\rho $ \emph{non-trivial}).
Then,
\begin{equation}
H_{\rho }^{q} ( {\cal G} , V ) = 0 \quad
\forall\, q \geq 0  \quad .
\label{4whitehead}
\end{equation}
If $q = 0$, the non-triviality of $\rho $ and the irreducibility imply that
$\rho ( {\cal G} ) \cdot v = 0 \ (v \in V)$ holds only for $v = 0$.
\end{lemma}

\begin{proof}
Since $\g$ is semi-simple, the Cartan-Killing metric $g_{ij}$ is
invertible, $g^{ij} g_{jk} = \delta^i_k$.
Let ${\tau}$ be the operator on the space of
$q$-cochains
${\tau}:C^q(\g,V)\to C^{q-1}(\g,V)$
defined by
\begin{equation}
( {\tau} \Omega )_{i_1 \dots i_{q-1}}^A =
g^{ij} \rho(X_i)^A_{.B} \Omega_{j i_1 \dots i_{q-1}}^B \quad.
\label{whitehead1}
\end{equation}
It is not difficult to check that on cochains the Laplacian-like
operator $(s\tau + \tau s)$ gives\footnote{
For instance, for a two-cochain eq. (\ref{whitehead2}) reads
\begin{equation}
\begin{array}{rl}
[(s\tau + \tau s)\Omega]^A_{ij} = &
g^{kl} \rho(X_i)^A_{.B} \rho(X_k)^B_{.C} \Omega^C_{l j}
  - g^{kl} \rho(X_j)^A_{.B} \rho(X_k)^B_{.C} \Omega^C_{l i}
  - g^{kl} \rho(X_k)^A_{.B} C_{ij}^m \Omega^B_{l m}
\\
+ & g^{kl} \rho(X_{k})^A_{.B} \rho(X_{l})^B_{.C} \Omega^C_{i j}
+ g^{kl} \rho(X_{k})^A_{.B} \rho(X_{i})^B_{.C} \Omega^C_{j l}
+ g^{kl} \rho(X_{k})^A_{.B} \rho(X_{j})^B_{.C} \Omega^C_{l i}
\\
- & g^{kl} \rho(X_{k})^A_{.B} C_{i j}^m \Omega^B_{m l}
- g^{kl} \rho(X_{k})^A_{.B} C_{l i}^m \Omega^B_{m j}
- g^{kl} \rho(X_{k})^A_{.B} C_{j l}^m \Omega^B_{m i}
\\
= & g^{kl} [\rho(X_i),\rho(X_k)]^A_{.B}\Omega^B_{l j}
- g^{kl} [\rho(X_j),\rho(X_k)]^A_{.B}\Omega^B_{l i}
+ I_2(\rho)^A_{.B} \Omega^B_{i j}
\\
- & g^{kl} \rho(X_{k})^A_{.B} C_{l i}^m \Omega^B_{m j}
-g^{kl} \rho(X_{k})^A_{.B} C_{j l}^m \Omega^B_{m i}
= I_2(\rho)^A_{.B} \Omega^B_{i j}\quad.
\end{array}
\end{equation}
}
\begin{equation}
[(s {\tau} + {\tau} s) \Omega] ^A _{i_1 \dots i_q} =
\Omega_{i_1 \dots i_q} ^B I_2(\rho)^A_{.B} \quad,
\label{whitehead2}
\end{equation}
where $I_2(\rho)^A_{.B}= g^{ij}(\rho(X_i) \rho(X_j))^A_{.B}$ is the
quadratic Casimir operator in the representation $\rho$.
By Schur's lemma it is proportional to the unit matrix.
Hence, applying (\ref{whitehead2}) to $\Omega\in Z^q_\rho(\g,V)$ we find
\begin{equation}
s {\tau} \Omega = \Omega I_2(\rho)
\; \Rightarrow\; s( {\tau} \Omega I_2(\rho)^{-1} ) = \Omega\quad.
\label{whitehead3}
\end{equation}
Thus, $\Omega$ is the coboundary
generated by the cochain ${\tau} \Omega I_2(\rho)^{-1} \in C^{q-1}_\rho(\g,V)$,
\emph{q.e.d.}
\end{proof}

For semisimple algebras and $\rho=0$ we also have $H_0^1=0$ and $H_0^2=0$, but
already $H_0^3\ne 0$.

\subsection{Lie algebra cohomology \emph{\`a la} BRST}
\label{sec5}

In many physical applications it is convenient to introduce the so-called BRST
operator (for Becchi, Rouet, Stora and Tyutin) acting on the space of BRST
cochains.
To this aim let us introduce anticommuting, `odd' objects (in physics they
correspond to the \emph{ghosts})
\begin{equation}
c^i c^j = - c^j c^i \quad,\quad i,j=1,\dots,\mbox{dim}\,\g\quad.
\label{brst1}
\end{equation}
The operator $\mathfrak{s}$ defined by
\begin{equation}
\mathfrak{s} := {1\over 2} C_{ij}^k c^j c^i {\partial\over \partial c^k}
\label{brst2}
\end{equation}
acts on the ghosts as the exterior derivative $d$ acts on LI one-forms
($\mathfrak{s} c^k = - 1/2 C_{ij}^k c^i c^j$,
cf. (\ref{3MC})) and, as $d$, is nilpotent, $\mathfrak{s}^2=0$.
For the cohomology associated with a non-trivial action $\rho$ of $\g$ on $V$
we introduce the BRST $\tilde s$ operator
\begin{equation}
\tilde s :=
c^i \rho(X_i) + {1\over 2} C_{ij}^k c^j c^i {\partial\over \partial c^k}
\quad.
\label{brst4}
\end{equation}

\begin{proposition}
\label{prop5.1}
The BRST operator $\tilde s$ is nilpotent $\tilde s^2 = 0$.
\end{proposition}

\begin{proof}
First, we rewrite $\tilde s$ as
\begin{equation}
{\tilde s}=c^i N_{(i)} \quad ,\quad
N_{(i)}=\rho (X_{i})+ {1\over 2} C^k_{ji} c^j {\partial\over \partial c^k}
\equiv N_{(i)}^1 +{1\over 2} N_{(i)}^2\quad .
\label{brst6}
\end{equation}
The operator $N_{(i)}$ has two different pieces $N^1$ and $N^2$, each of
them carrying a representation of \g\ so that
$[N_{(i)},N_{(j)}]=C^k_{ij} (N_{(k)}^1 + {1\over 4}N_{(k)}^2)$.
Thus,
\begin{equation}
\begin{array}{l}
{\tilde s}^2
= \displaystyle
c^i N_{(i)} c^j N_{(j)} = {1\over 2} c^i c^j [N_{(i)},N_{(j)}]+
c^i(N_{(i)}.c^j)N_{(j)}
\\[0.3cm]
\displaystyle
\ \
= {1\over 2} c^i c^j C^k_{ij} (N_{(k)}^1 + {1\over 4}N_{(k)}^2)+
{1\over 2} c^i c^j C^k_{ji} N_{(k)}=
{1\over 2} c^i c^j C^k_{ij} N_{(k)}^1
+ {1\over 2} c^i c^j C^k_{ji} N_{(k)}^1 = 0
\ ,
\end{array}
\label{brst8}
\end{equation}
by virtue of the anticommutativity of the $c$'s, and
using that $c^i c^j C^k_{ij} N_{(k)}^2=0$ and
$N_{(i)}.c^j = {1\over 2} c^k C^j_{ki}$.
Thus, on the `BRST-cochains'
\begin{equation}
\tilde \Omega^A_n = {1\over n!} \Omega_{i_1 \dots i_n} ^A c^{i_1}\dots c^{i_n}
\quad,
\label{brst9}
\end{equation}
the action of $\tilde s$ is the same as that of $s$ in
(\ref{4cobound}) and may be used to define the Lie algebra cohomology.
\end{proof}

\section{Symmetric polynomials and higher order cocycles}
\label{sec6}

\subsection{Symmetric invariant tensors and higher order Casimirs}
\label{sec6.1}

{}From now on, we shall restrict ourselves to simple Lie groups
and algebras; by virtue of Lemma~\ref{lem4.1}, only the $\rho=0$ case is
interesting.
The non-trivial cohomology groups are related to the primitive
symmetric invariant tensors
\cite{Rac:50,Gel:50,Kle:63,Gru.Rai:64,Bie:63,Per.Pop:68,Oku.Pat:83,Oku:82}
on \g, which in turn determine Casimir elements in the universal
enveloping algebra ${\cal U}({\cal G})$.

\begin{definition}
[{\normalfont\textit{Symmetric and invariant polynomials on \g}}{}]
\label{def6.1}
A symmetric polynomial on $\cal G$ is given by a symmetric covariant LI tensor.
It may be expressed as a LI covariant tensor on $G$,
$k=k_{i_1...i_m}\omega^{i_1}\otimes...\otimes\omega^{i_m}$
with symmetric constant coordinates $k_{i_1...i_m}$.
$k$ is said to be an invariant or ($ad$-invariant) symmetric polynomial if it
is also
right-invariant, \ie\ if $L_{X_l}k=0$ $\forall\, X_l\in\mathfrak{X}^L(G)$.
Indeed, using (\ref{3Lalpha}), we find that
\begin{equation}
L_{X_l} k=0\ \Rightarrow\
C^s_{li_1}k_{si_2...i_m}+C^s_{li_2}k_{i_1s...i_m}+\dots+
C^s_{li_m}k_{i_1...i_{m-1}s}
=0\quad .                                          \label{kinva}
\end{equation}
Since the coordinates of $k$ are given by
$k_{i_1\dots i_m}=   k(X_{i_1},\dots,X_{i_m})$,
eq. (\ref{kinva}) is equivalent to stating that $k$ is $ad$-invariant, \ie,
\begin{equation}
  k([X_l,X_{i_1}],\dots,X_{i_m})+k(X_{i_1},[X_l,X_{i_2}],\dots,X_{i_m})
 +\dots + k(X_{i_1},\dots,[X_l,X_{i_m}])=0              \label{adinfi}
\end{equation}
or, equivalently,
\begin{equation}
 k(Ad\,g\, X_{i_1},\dots,Ad\,g\, X_{i_m})=k(X_{i_1},\dots,X_{i_m})\quad,
                                                          \label{adfini}
\end{equation}
from which eq. (\ref{adinfi}) follows by taking the derivative
$\partial /\partial g ^l$ in $g=e$.
\end{definition}

The invariant symmetric polynomials just described can be used to construct
Casimir elements of the enveloping algebra ${\cal U}(\g)$ of $\g$ in the
following way

\begin{proposition}
\label{prop6.1}
Let $k$ be a symmetric invariant tensor.
Then $k^{i_1\dots i_m}X_{i_1}\dots X_{i_m}$
(coordinate indices of $k$ raised using the Killing metric),
is a Casimir of order $m$,
\ie\ $[k^{i_1\dots i_m}X_{i_1}\dots X_{i_m},
Y]=0$ $\forall\, Y\in\g$.
\end{proposition}

\begin{proof}
\begin{equation}
\eqalign{
[k^{i_1\dots i_m}X_{i_1}\dots X_{i_m},X_s]
=&
\sum^m_{j=1}k^{i_1\dots i_m}X_{i_1}\dots [X_{i_j},X_s]\dots X_{i_m}
\cr
=&
\sum^m_{j=1}k^{i_1\dots i_m}X_{i_1}\dots C^t_{i_j s}X_t\dots X_{i_m}=0
\cr}
\label{casimir}
\end{equation}
by (\ref{kinva}), \emph{q.e.d.}
\end{proof}

A well-known way of obtaining symmetric ($ad$-)invariant polynomials
(used \eg, in the construction of characteristic classes) is given by

\begin{proposition}
\label{prop6.2}
Let $X_i$ denote now a representation of $\g$. Then, the symmetrized trace
\begin{equation}
     k_{i_1\dots i_m}=\hbox{sTr}(X_{i_1}\dots X_{i_m})   \label{symtra}
\end{equation}
defines a symmetric invariant polynomial.
\end{proposition}

\begin{proof}
$k$ is symmetric by construction and the $ad$-invariance is obvious since
$Adg\,X:=gXg^{-1}$, \emph{q.e.d.}
\end{proof}
The simplest illustration of (\ref{symtra}) is the Killing tensor for a simple
Lie algebra $\g$, $k_{ij}=\hbox{Tr}(ad\, X_i\, ad\, X_j)$;
its associated Casimir is the second order Casimir $I_2$.

\begin{example}
\label{example6.2}
Let $\g=su(n)$, $n\ge2$, and let $X_i$ be (hermitian) matrices in the
defining representation. Then
\begin{equation}
     \hbox{sTr}(X_iX_jX_k)\propto 2\hbox{Tr}(\{ X_i,X_j\} X_k)=d_{ijk}\quad,
\label{deij}
\end{equation}
using that, for the $su(n)$ algebra, $\{ X_i,X_j\}=c\delta_{ij}
+d_{ijl}X_l$, $\hbox{Tr}(X_k)=0$ and $\hbox{Tr}(X_iX_j)={1\over 2}
\delta_{ij}$.
This third order polynomial leads to the Casimir $I_3$;
for $su(2)$ only $k_{ij}$ and $I_2$ exist.
\end{example}

\begin{example}
\label{example6.3}
In the case $\g=su(n)$, $n\ge 4$, we have a fourth order polynomial
\begin{equation}
 \hbox{sTr}(X_{i_1}X_{i_2}X_{i_3}X_{i_4})\propto d_{(i_1i_2l}d_{li_3)i_4}
+2 c\delta_{(i_1i_2}\delta_{i_3)i_4} \quad,      \label{kcuatro}
\end{equation}
where $(\ )$ indicates symmetrization.
The first term leads to a fourth order Casimir $I_4$ whereas the second
one includes (see \cite{Azc.Mac.Mou.Bue:97}) a term in $I_2^2$.
\end{example}

Eq. (\ref{kcuatro}) deserves a comment. The first part
$d_{(i_1i_2l}d_{li_3)i_4}$ generalizes easily to higher $n$ by nesting more
$d$'s, leading to the Klein \cite{Kle:63} form of the $su(n)$ Casimirs.
The second part includes a term that is the product of Casimirs of order two:
it is not \emph{primitive}.

\begin{definition}
[{\normalfont\textit{Primitive symmetric invariant polynomials}}{}]
\label{def6.2}
A symmetric invariant polynomial $k_{i_1\dots i_m}$ on $\g$ is called
primitive if it is not of the form
\begin{equation}
   k_{i_1\dots i_m}=k^{(p)}_{(i_1\dots i_p}k^{(q)}_{i_{p+1}\dots i_m)}
\ ,\quad p+q=m\quad,
\label{primitive}
\end{equation}
where $k^{(p)}$ and $k^{(q)}$ are two lower order symmetric invariant
polynomials.
\end{definition}

Of course, we could also have considered eq. (\ref{kcuatro}) for $su(3)$, but
then it would
not have led to a fourth-order primitive polynomial, since $su(3)$
is a rank 2 algebra.
Indeed, $d_{(i_1i_2l}d_{li_3)i_4}$ is not primitive for $su(3)$
and can be written in terms of $\delta_{i_1i_2}$ as in (\ref{primitive}) (see,
\eg, \cite{Sud:90}; see also \cite{Azc.Mac.Mou.Bue:97} and references therein).
In general, for a simple algebra of rank $l$ there are $l$ invariant primitive
polynomials and Casimirs
\cite{Rac:50,Gel:50,Kle:63,Gru.Rai:64,Bie:63,Per.Pop:68,Oku.Pat:83,Oku:82}
and, as we shall show now, $l$ primitive Lie algebra cohomology cocycles.

\subsection{Cocycles from invariant polynomials}
\label{sec6.2}

We make now explicit the connection between the invariant
polynomials and the non-trivial cocycles of a simple Lie algebra $\g$. To do
this we may use the particular case of $\g=su(n)$ as a guide.
On the manifold of the group $SU(n)$ one can construct the \emph{odd} $q$-form
\begin{equation}
    \Omega=\frac{1}{q!} \hbox{Tr}(\theta\wedge \mathop{\cdots}\limits^q
     \wedge\theta)
     \quad ,                                            \label{HOCa}
\end{equation}
where $\theta=\omega^i X_i$ and we take
$\{X_i\}$ in the defining representation;
$q$ has to be odd since otherwise $\Omega$ would be zero (by virtue of the
cyclic property of the trace and the anticommutativity of one-forms).

\begin{proposition}
\label{prop6.3}
The LI odd form $\Omega$ on $G$ in (\ref{HOCa}) is a non-trivial
(CE) Lie algebra cohomology cocycle.
\end{proposition}

\begin{proof}
Since $\Omega$ is LI by construction, it is sufficient to
show that $\Omega$ is closed and that it is not the differential of another
LI form (\ie\ it is not a coboundary).
By using (\ref{3canonMC}) we get
\begin{equation}
       d\Omega=-\frac{1}{(q-1)!}\hbox{Tr}(\theta\wedge
\mathop{\cdots}\limits^{q+1}
       \wedge\theta)=0\quad,            \label{HOCb}
\end{equation}
since $q+1$ is even.
Suppose now that $\Omega=d\Omega_{q-1}$, with $\Omega_{q-1}$ LI.
Then $\Omega_{q-1}$ would be of the form (\ref{HOCa}) and hence
zero because $q-1$ is also even, \emph{q.e.d.}
\end{proof}

All non-trivial $q$-cocycles in $H_0^q(su(n),\R)$ are of the form (\ref{HOCa}).
The fact that they are closed and non-exact ($SU(n)$ is compact) allows
us to use them to construct Wess-Zumino-Witten \cite{Wit:84,Wit:83} terms
on the group manifold (see also \cite{Azc.Izq.Mac:90}).

Let us set $q=2m-1$. The form $\Omega$ expressed in coordinates is
\begin{equation}
\eqalign{\Omega&=\frac{1}{q!}\hbox{Tr}(X_{i_1}\dots X_{i_{2m-1}})
\omega^{i_1}\wedge\dots\wedge\omega^{i_{2m-1}}
\cr &
\propto \hbox{Tr}([X_{i_1},X_{i_2}][X_{i_3},X_{i_4}]\dots
[X_{i_{2m-3}},X_{i_{2m-2}}]X_{i_{2m-1}})
\omega^{i_1}\wedge\dots\wedge\omega^{i_{2m-1}}
\cr &
=\hbox{Tr}(X_{l_1}\dots X_{l_{m-1}}X_\sigma)C^{l_1}_{i_1i_2}
\dots C^{l_{m-1}}_{i_{2m-3}i_{2m-2}}
\omega^{i_1}\wedge\dots\wedge\omega^{i_{2m-2}}\wedge\omega^\sigma
      \quad . \cr}                               \label{HOCc}
\end{equation}
We see here how
the order $m$ symmetric (there is symmetry in ${l_1}\dots l_{m-1}$ because of
the $\omega^{i}$'s) invariant polynomial $\hbox{Tr}(X_{l-1}\dots
X_{l_{m-1}}X_\sigma)$ appears in this context. Conversely, the following
statement holds

\begin{proposition}
\label{prop6.4}
Let $k_{i_1\dots i_m}$ be a symmetric invariant polynomial. Then, the
polynomial
\begin{equation}
    \Omega_{\rho i_2\dots i_{2m-2}\sigma}=C^{l_1}_{j_2j_3}\dots
    C^{l_{m-1}}_{j_{2m-2}\sigma}k_{\rho l_1\dots l_{m-1}}
     \varepsilon^{j_2\dots j_{2m-2}}_{i_2\dots i_{2m-2}}  \label{HOCe}
\end{equation}
is skew-symmetric and defines the closed form (cocycle)
\begin{equation}
\Omega=\frac{1}{(2m-1)!}\Omega_{\rho i_2\dots i_{2m-2}\sigma}\omega^\rho\wedge
    \omega^{i_2}\wedge\dots\wedge\omega^{i_{2m-2}}\wedge\omega^\sigma
\quad.
\label{HOCd}
\end{equation}
\end{proposition}

\begin{proof}
To check the complete skew-symmetry of
$\Omega_{\rho i_2\dots i_{2m-2}\sigma}$ in (\ref{HOCe}), it
is sufficient, due to the $\varepsilon$, to show the antisymmetry in
$\rho$ and $\sigma$.
This is done by using the invariance of $k$ (\ref{kinva})
and the symmetry properties of $k$ and $\varepsilon$ to rewrite
$\Omega_{\rho i_2\dots i_{2m-2}\sigma}$ as the sum of two terms. The first one,
\begin{equation}
       \eqalign{&\sum^{m-2}_{s=1}\varepsilon^{j_2\dots j_{2s}j_{2s+1}
j_{2m-2}j_{2s+2}\dots j_{2m-3}}_{i_2\dots i_{2m-2}}
    k_{\rho l_1\dots l_{s-1}l_{m-1}l_s\dots l_{m-2}\sigma}\cr
      &C^{l_1}_{j_2j_3}\dots C^{l_s}_{j_{2s}j_{2s+1}}
     C^{l_{m-1}}_{l_sj_{2m-2}}C^{l_{s+1}}_{j_{2s+2}j_{2s+3}}\dots
      C^{l_{m-2}}_{j_{2m-4}j_{2m-3}} \cr}          \label{HOCf}
\end{equation}
vanishes due to the Jacobi identity in (\ref{3jacobi}), and the second one is
\begin{equation}
    \eqalign{\Omega_{\rho i_2\dots i_{2m-2}\sigma}=
     -\varepsilon^{j_2\dots j_{2m-2}}_{i_2\dots i_{2m-2}}k_{\sigma l_1\dots
 l_{m-1}} C^{l_1}_{j_2j_3}\dots C^{l_{m-1}}_{j_{2m-2}\rho}
    =-\Omega_{\sigma i_2\dots i_{2m-2}\rho}\quad .\cr}  \label{HOCg}
\end{equation}
To show that $d\Omega=0$ we make use of the fact that any bi-invariant form
(\ie, a form that is both LI and RI) is closed (see, \eg, \cite{Azc.Izq:95}).
Since
$\Omega$ is LI by construction, we only need to prove its
right-invariance, but
\begin{equation}
\Omega\propto
\mbox{Tr}(\theta \wedge \mathop{\cdots}\limits^{2m-1} \wedge \theta)
\label{HOCh}
\end{equation}
is obviously RI since $R_g^*\theta = Ad g^{-1} \theta$, \emph{q.e.d.}
\end{proof}

Without discussing the origin of the invariant polynomials for the
different groups
\cite{Rac:50,Gel:50,Kle:63,Gru.Rai:64,Bie:63,Per.Pop:68,Oku.Pat:83,Oku:82,
Azc.Mac.Mou.Bue:97},
we may conclude that to each symmetric primitive invariant
polynomial of order $m$ we can associate a Lie algebra cohomology
$(2m-1)$-cocycle (see \cite{Azc.Mac.Mou.Bue:97} for practical details).
The question that immediately arises is whether this construction may be
extended since, from a set of $l$ primitive invariant
polynomials, we can obtain an arbitrary number of non-primitive polynomials
(see eq.~(\ref{primitive})).
This question is answered negatively by Proposition~\ref{prop6.5} and
Corollary~\ref{cor6.1} below.

\begin{proposition}
\label{prop6.5}
Let $k_{i_1\dots i_m}$ be a symmetric $G$-invariant polynomial. Then,
\begin{equation}
\epsilon^{j_1 \dots j_{2m}}_{i_1\dots i_{2m}}
C^{l_1}_{j_1 j_2}\ldots C^{l_m}_{j_{2m-1} j_{2m}}
k_{l_1\dots l_m}=0 \quad.
\label{simplelemma}
\end{equation}
\end{proposition}

\begin{proof}
By replacing $C^{l_m}_{j_{2m-1} j_{2m}}k_{l_1\dots lm}$ in the $l.h.s$ of
(\ref{simplelemma}) by the other terms in (\ref{kinva}) we get
\begin{equation}
\epsilon^{j_1 \dots j_{2m}}_{i_1 \dots i_{2m}}
C^{l_1}_{j_1 j_2}\ldots C^{l_{m-1}}_{j_{2m-3} j_{2m-2}}
(\sum_{s=1}^{m-1}C^{k}_{j_{2m-1} l_s} k_{l_1\ldots l_{s-1} k l_{s+1}\ldots
l_{m-1}\,j_{2m}})
\quad,
\end{equation}
which is zero due to the JI, \emph{q.e.d.}
\end{proof}

\begin{corollary}
\label{cor6.1}
Let $k$ be a non-primitive symmetric invariant polynomial (\ref{primitive}),
Then the $(2m-1)$-cocycle $\Omega$ associated to it (\ref{HOCd}) is zero.
\end{corollary}

Thus, to a \emph{primitive} symmetric $m$-polynomial it is possible
to associate uniquely a Lie algebra $(2m-1)$-cocycle.
Conversely, we also have the following

\begin{proposition}
\label{prop6.6}
Let $\Omega^{(2m-1)}$ be a primitive cocycle.
The $l$ polynomials $t^{(m)}$ given by
\begin{equation}
{t}^{i_1 \dots i_{m}}=[\Omega^{(2m-1)}]^{j_1 \dots j_{2m-2} i_m}
C^{i_1}_{j_1 j_2} \dots C^{i_{m-1}}_{j_{2m-3} j_{2m-2}}
\label{seIIxii}
\end{equation}
are invariant, symmetric and primitive (see
\cite[Lemma 3.2]{Azc.Mac.Mou.Bue:97}).
\end{proposition}

This converse proposition relates the cocycles of the Lie algebra
cohomology to Casimirs in the enveloping algebra ${\cal U}(\g)$.
The polynomials in (\ref{seIIxii}) have certain advantages (for instance, they
have all traces equal to zero) \cite{Azc.Mac.Mou.Bue:97} over other more
conventional ones such as \eg, those in (\ref{symtra}).

\subsection{The case of simple compact groups}
\label{sec7}

We have seen that the
Lie algebra cocycles may be expressed in terms of LI forms on the group
manifold $G$ (Sec.~\ref{sec4.2}).
For compact groups, the CE cohomology can be identified
(see, \eg\ \cite{Che.Eil:48}) with the de Rham cohomology:

\begin{proposition}
\label{prop7.1}
Let $G$ be a compact  and connected Lie group. Every de Rham cohomology class
on $G$ contains one and only one bi-invariant form. The bi-invariant forms span
a ring isomorphic to $H_{DR}(G)$.
\end{proposition}

The equivalence of the Lie algebra (CE) cohomology and the de Rham cohomology
is specially interesting because, since all primitive cocycles are odd,
compact groups behave as products of odd spheres
from the point of view of real homology.
This leads to a number of simple an elegant formulae concerning the Poincar\'e
polynomials, Betti numbers, etc.
We conclude by giving a table (table~\ref{table7.1})
which summarizes many of these results.
Details on the topological properties of Lie groups may be found in
\cite{Car:36,Pon:35,Hop:41,Sam:52,Bor:65,Bot:79,Boy:93};
for book references see \cite{Wey:46,Hod:41,Gre.Hal.Van:76,Azc.Izq:95}.

\begin{table}[H]
$$
\begin{array}{cccc}
\g & \mbox{dim}\,\g & \mbox{order\ of\ invariants\ and\ Casimirs}
& \mbox{order\ of\ $\g$-cocycles}
\\
\hline
\hline
A_l & (l+1)^2 -1\ [l>1] & 2,3,\dots,l+1         & 3,5,\dots,2l+1
\\
B_l & l(2l+1)\ [l>2]    & 2,4,\dots,2l          & 3,7,\dots,4l-1
\\
C_l & l(2l+1)\ [l>3]    & 2,4,\dots,2l          & 3,7,\dots,4l-1
\\
D_l & l(2l-1)\ [l>4]    & 2,4,\dots,2l-2,l      & 3,7,\dots,4l-5,2l-1
\\
G_2 & 14                & 2,6                   & 3,11
\\
F_4 & 52                & 2,6,8,12              & 3,11,15,23
\\
E_6 & 78                & 2,5,6,8,9,12          & 3,9,11,15,17,23
\\
E_7 & 133               & 2,6,8,10,12,14,18     & 3,11,15,19,23,27,35
\\
E_8 & 248               & 2,8,12,14,18,20,24,30 & 3,15,23,27,35,39,47,59
\end{array}
$$
\caption{Order of the primitive invariant polynomials and associated cocycles
for all the simple Lie algebras.}
\label{table7.1}
\end{table}

\section{Higher order simple and SH Lie algebras}
\label{sec8}

We present here a construction for which the previous cohomology notions
play a crucial role, namely the construction of higher order Lie algebras.
Recall that ordinary Lie algebras are defined as vector spaces endowed with
the Lie bracket, which obeys the JI.
If the Lie algebra is simple
$\omega_{ij\rho}=k_{\rho\sigma}C^\sigma_{ij}$ is the non-trivial three-cocycle
associated with the Cartan-Killing metric, given by the structure constant
themselves (see (\ref{HOCe})).
The question arises as to whether higher order cocycles (and therefore Casimirs
of order higher than two) can be used to define the structure constants of a
higher order bracket. Given the odd-dimension of the cocycles, these
multibrackets will involve an even number of Lie algebra elements.
Since we already have matrix
realizations of the simple Lie algebras, let us use them to construct the
higher order brackets.
Consider the case of $su(n)$, $n>2$ and a
four-bracket. Let $X_i$ be the matrices of the defining representation.
Since the
bracket has to be totally skew-symmetric, a sensible definition for it is
\begin{equation}
 [X_{i_1},X_{i_2},X_{i_3},X_{i_4}]:=\varepsilon^{j_1j_2j_3j_4}_{i_1i_2i_3i_4}
  X_{j_1}X_{j_2}X_{j_3}X_{j_4}   \quad .                    \label{multia}
\end{equation}
This four-bracket generalizes the ordinary \hbox{(two-)} bracket
$[X_{i_1},X_{i_2}]=\varepsilon^{j_1j_2}_{i_1i_2}X_{j_1}X_{j_2}$.
By using the skew-symmetry in $j_1\dots j_4$, we may
rewrite (\ref{multia}) in terms of commutators as
\begin{equation}
  \eqalign{ [X_{i_1},X_{i_2},X_{i_3},X_{i_4}]&= \frac{1}{2^2}
 \varepsilon^{j_1j_2j_3j_4}_{i_1i_2i_3i_4}[X_{j_1},X_{j_2}][X_{j_3},X_{j_4}]
 =\frac{1}{2^2}\varepsilon^{j_1j_2j_3j_4}_{i_1i_2i_3i_4}
   C^{l_1}_{j_1j_2}C^{l_2}_{j_3j_4}X_{l_1}X_{l_2}\cr
    &=\frac{1}{2^2}\varepsilon^{j_1j_2j_3j_4}_{i_1i_2i_3i_4}
        C^{l_1}_{j_1j_2}C^{l_2}_{j_3j_4}\frac{1}{2}
     ({d_{l_1l_2}}^\sigma_{.} X_\sigma+c\delta_{l_1l_2})\cr
     &=\frac{1}{2^3}\varepsilon^{j_1j_2j_3j_4}_{i_1i_2i_3i_4}
     C^{l_1}_{j_1j_2}C^{l_2}_{j_3j_4}{d_{l_1l_2}}^\sigma_{.} X_\sigma
  ={\omega_{i_1\dots i_4}}^\sigma_{.} X_\sigma\quad ,\cr} \label{multib}
\end{equation}
where in going from the first line to the second we have used that the factor
multiplying $X_{l_1}X_{l_2}$ is symmetric in $l_1,l_2$, so that we can
replace $X_{l_1}X_{l_2}$ by ${1\over 2}\{X_{l_1},X_{l_2}\}$ and then
write it in terms of the $d$'s. The contribution of the term proportional to
$c$ vanishes due to the JI. Thus,
the structure constants of the four-bracket are given by the $5$-cocycle
corresponding to the primitive polynomial $d_{ijk}$. These
reasonings can be generalized to higher order brackets and to the
other simple algebras. This motivates the following

\begin{definition}[{\normalfont\textit{Higher order bracket}}{}]
\label{def8.1}
Let $X_i$ be arbitrary associative operators.
The corresponding higher order bracket or
multibracket of order $n$ is defined by \cite{Azc.Bue:97}
\begin{equation}
     [X_1,\dots,X_n]: =\sum_{\sigma\in S_n}(-1)^{\pi(\sigma)}
  X_{i_{\sigma(1)}}\dots X_{i_{\sigma(n)}}             \quad .  \label{multic}
\end{equation}
\end{definition}
The bracket (\ref{multic}) obviously satisfies the JI when $n=2$.
In the general case, the situation depends on whether $n$ is even or odd, as
stated by

\begin{proposition}
\label{prop8.1}
For $n$ even, the $n$-bracket (\ref{multic}) satisfies the generalized Jacobi
identity (GJI) \cite{Azc.Bue:97}
\begin{equation}
    \sum_{\sigma\in S_{2n-1}}(-1)^{\pi(\sigma)}\left[ [X_{\sigma(1)},\dots,
     X_{\sigma(n)}],X_{\sigma(n+1)},\dots ,X_{\sigma(2n-1)}\right] =0
                                                 \label{multie}
\quad;
\end{equation}
for $n$ odd, the l.h.s. of (\ref{multie}) is
proportional to $[X_1,\dots , X_{2n-1}]$.
\end{proposition}

\begin{proof}
In terms of the Levi-Civita symbol, the l.h.s. of (\ref{multie}) reads
\begin{equation}
\varepsilon^{j_1\dots j_{2n-1}}_{i_1\dots i_{2n-1}}
\varepsilon^{l_1\dots l_n}_{j_1\dots j_n} [X_{l_1}\cdots X_{l_n},X_{j_{n+1}},
\dots ,X_{j_{2n-1}}]
\quad.
\label{laultima}
\end{equation}
Notice that the product $X_{l_1}\cdots X_{l_n}$ is a single entry in the
$n$-bracket $[X_{l_1}\cdots X_{l_n},X_{j_{n+1}},\dots ,\allowbreak
X_{j_{2n-1}}]$.
Since the $n$ entries in this bracket are also antisymmetrized, eq.
(\ref{laultima}) is equal to
\begin{equation}
\eqalign{
& n!
\varepsilon^{l_1\dots l_n j_{n+1}\dots j_{2n-1}}_{i_1\dots\dots\dots i_{2n-1}}
   \varepsilon^{l_{n+1}\dots l_{2n-1}}_{j_{n+1}\dots j_{2n-1}}
   \sum^{n-1}_{s=0}(-1)^s X_{l_{n+1}}\cdots X_{l_{n+s}}X_{l_1}\cdots X_{l_n}
     X_{l_{n+1+s}}\cdots X_{l_{2n-1}}
\cr
= & n!(n-1)!\varepsilon^{l_1\dots l_{2n-1}}_{i_1\dots i_{2n-1}}
    X_{l_1}\cdots X_{l_{2n-1}}\sum^{n-1}_{s=0}(-1)^s(-1)^{ns}
\cr
= &
    n!(n-1)! [X_{i_1},\dots ,X_{i_{2n-1}}]\sum^{n-1}_{s=0}(-1)^{s(n+1)}\quad,
\cr}
\label{multif}
\end{equation}
where we have used the skew-symmetry of $\varepsilon$ to relocate the block
$X_{l_1}\cdots X_{l_n}$ in the second equality.
Thus, the l.h.s. of (\ref{multie}) is proportional to a multibracket of order
$(2n-1)$ times a sum, which for even $n$ vanishes and for odd $n$ is equal to
$n$, \emph{q.e.d.}
\end{proof}

In view of the above result, we introduce the following definition
\cite{Azc.Bue:97}

\begin{definition}
[{\normalfont\textit{Higher order Lie algebra}}{}]
\label{def8.2}
An order $n$ ($n$ even) generalized Lie algebra is a vector space $V$ of
elements $X\in V$ endowed with a fully skew-symmetric bracket
$V\times \mathop{\cdots}\limits^n \times V\rightarrow V$,
$(X_1,\dots ,X_n)\mapsto
[X_1,\dots ,X_n] \in V$ such that the GJI
(\ref{multie}) is fulfilled.
\end{definition}

Consequently, a finite-dimensional Lie algebra of order $n=2p$, generated by
the elements $\{ X_i\}_{i=1,\dots,r}$ will be defined by an equation of the
form
\begin{equation}
 [X_{i_1},\dots ,X_{i_{2p}}]={C_{i_1\dots i_{2p}}}^j X_j\quad ,\label{multig}
\end{equation}
where ${C_{i_1\dots i_{2p}}}^j$ are the generalized structure constants. An
example of this is provided by the construction given in (\ref{multib}), where
the bracket is defined as in (\ref{multic}) and the structure constants are
$(2p+1)$-cocycles of the simple Lie algebra used, $\Omega_{i_1\dots i_{2p}
\sigma}$. Writing now the GJI (\ref{multie}) in
terms of the $\Omega$'s, the following equation is obtained
\begin{equation}
   \varepsilon^{j_1\dots j_{4p-1}}_{i_1\dots i_{4p-1}}
      {\Omega_{j_1\dots j_{2p}}}^\sigma\Omega_{\sigma j_{2p+1}\dots j_{4p-1}
    \rho}=0 \quad .                                     \label{multih}
\end{equation}
This equation is known to hold due to Proposition~\ref{prop8.1}
and a generalization
of the argument given in (\ref{multib}), which in fact provides the proof of

\begin{theorem}
[{\normalfont\textit{Classification theorem for higher-order simple Lie
algebras}}{}]
\label{th8.1}
Given a simple algebra $\g$ of rank $l$, there are $l-1$
$(2m_i-2)$-higher-order simple Lie algebras associated with $\g$. They are
given by the $l-1$ Lie algebra cocycles of order $2m_i-1>3$ which may be
obtained from the $l-1$ symmetric invariant polynomials on $\g$ of order
$m_i>m_1=2$. The $m_1=2$ case (Killing metric) reproduces the original simple
Lie algebra $\g$; for the other $l-1$ cases, the skew-symmetric
$(2m_i-2)$-commutators define an element of $\g$ by means of the
$(2m_i-1)$-cocycles. These higher-order structure constants (as the ordinary
structure constants with all the indices written down) are fully antisymmetric
cocycles and satisfy the GJI.
\end{theorem}

\begin{proposition}
[{\normalfont\textit{Mixed order generalized Jacobi identity}}{}]
Let $m,n$ be even.
We introduce the mixed order generalized Jacobi identity for even order
multibrackets by
\begin{equation}
       \varepsilon^{j_1\dots j_{n+m-1}}\left[ [X_{j_1},\dots ,X_{j_n}],
   \dots , X_{j_{n+m-1}}\right]=0 \quad .                    \label{multii}
\end{equation}
\end{proposition}

\begin{proof}
Following the same reasonings of Proposition~\ref{prop8.1},
\begin{equation}
  \eqalign{
&  \varepsilon^{j_1\dots j_{n+m-1}}_{i_1\dots i_{n+m-1}}
  \varepsilon^{l_1\dots l_n}_{j_1\dots j_n} [X_{l_1}\cdots X_{l_n},X_{j_{n+1}},
   \dots ,X_{j_{n+m-1}}]
\cr
&\quad = n!
\varepsilon^{l_1\dots l_n j_{n+1}\dots j_{n+m-1}}_{i_1\dots\dots\dots
i_{n+m-1}}
\varepsilon^{l_{n+1}\dots l_{n+m-1}}_{j_{n+1}\dots j_{n+m-1}}
\sum^{m-1}_{s=0}(-1)^s X_{l_{n+1}}\cdots X_{l_{n+s}} X_{l_1}\cdots X_{l_n}
     X_{l_{n+1+s}}\cdots X_{l_{n+m-1}}
\cr
&\quad =n!(m-1)!\varepsilon^{l_1\dots l_{n+m-1}}_{i_1\dots i_{n+m-1}}
    X_{l_1}\cdots X_{l_{n+m-1}}\sum^{m-1}_{s=0}(-1)^s(-1)^{ns}
\cr
&\quad =
n!(m-1)! [X_{i_1},\dots,X_{i_{n+m-1}}]
\sum^{m-1}_{s=0}(-1)^{(n+1)s}\quad,
\cr}
\label{multifnew}
\end{equation}
which is zero for $n$ and $m$ even.
In contrast, if $n$ and/or $m$ are odd the sum
$\displaystyle \sum^{m-1}_{s=0}(-1)^{(n+1)s}$
is different from zero ($m$ if $n$ is odd and 1 if $n$ is even).
In this case, the l.h.s. of (\ref{multii}) is proportional to the
$(n+m-1)$-commutator $[X_{i_1},\dots,X_{i_{n+m-1}}]$, \emph{q.e.d.}
\end{proof}

In particular, if $n$ and $m$ are the orders of higher order algebras, the
identity (\ref{multii}) leads to (cf. (\ref{multih}))
\begin{equation}
  \varepsilon^{i_1\dots i_{n+m-1}}{\Omega_{i_1\dots i_n}}^\sigma
       \Omega_{\sigma i_{n+1}\dots i_{n+m-1}\rho}=0 \quad .\label{multij}
\end{equation}
For $n=2$ and $[X_i,X_j]=C_{ij}^k X_k$,
$[X_{i_1},\dots,X_{i_m}]={\Omega_{i_1\dots i_m}}^k X_k$ eq. (\ref{multij})
gives
\begin{equation}
  \varepsilon^{i_1\dots i_{m+1}}C_{i_1 i_2}^\sigma
       \Omega_{\sigma i_{3}\dots i_{m+1}\rho}=0 \quad,\label{multija}
\end{equation}
which implies that $\Omega_{i_1 \dots i_{m+1}}$ is a cocycle, \ie,
\begin{equation}
  \varepsilon^{i_1\dots i_{m+2}} C_{i_1 i_2}^\sigma
       \Omega_{\sigma i_{3}\dots i_{m+1} i_{m+2}}=0 \quad.
\label{multijb}
\end{equation}
Expression (\ref{multijb}) follows from (\ref{multija}), simply
antisymmetrizing the index $\rho$.

\subsection{Multibrackets and coderivations}
\label{sec9}
Higher-order brackets can be used to
generalize the ordinary coderivation of multivectors.

\begin{definition}
\label{def9.1}
Let $\{ X_i\}$ be a basis of $\g$ given in terms of
LIVF on $G$, and $\wedge^*(\g)$ the exterior algebra of multivectors
generated by them
($X_1\wedge\dots\wedge X_{q}\equiv\varepsilon_{1\dots q}^{i_1\dots i_q}
X_{i_1}\otimes\dots\otimes X_{i_q}$).
The exterior coderivation
$\partial:\wedge^q\rightarrow\wedge^{q-1}$ is given by
\begin{equation}
       \partial(X_1\wedge\dots\wedge X_q)=\sum^q_{l=1\atop l<k}(-1)^{l+k+1}
   [X_l,X_k]\wedge X_1\wedge\dots\wedge{\hat X_l}\wedge\dots
\wedge{\hat X_k}\wedge\dots\wedge
     X_q\quad .                                                \label{codea}
\end{equation}

This definition is analogous to that of the exterior derivative $d$, as given
by (\ref{prelk}) with its first term missing when one considers left-invariant
forms (eq. (\ref{3diff})). As $d$, $\partial$ is nilpotent, $\partial^2=0$,
due to the JI for the commutator.
\end{definition}

In order to generalize (\ref{codea}), let us note that
$\partial(X_1\wedge X_2)=[X_1,X_2]$,
so that (\ref{codea}) can be interpreted as a formula that gives the action
of $\partial$ on a $q$-vector in terms of that on a bivector.
For this reason we may write $\partial_2$ for $\partial$ above.
It is then natural to introduce an operator
$\partial_s$ that on a $s$-vector
gives the multicommutator of order $s$.
On an $n$-multivector its action is given by
\begin{definition}[{\normalfont\textit{Coderivation $\partial_s$}}{}]
\label{def9.2}
The general coderivation $\partial_s$ of degree $-(s-1)$ ($s$ even)
$\partial_s:\wedge^n(G)\rightarrow \wedge^{n-(s-1)} (G)$
is defined by
\begin{equation}
\eqalign{
&
\partial_s(X_1\wedge\dots\wedge X_n):=\frac{1}{s!}\frac{1}{(n-s)!}
\varepsilon^{i_1\dots i_n}_{1\dots n}\partial_s(X_{i_1}\wedge\dots\wedge
X_{i_s})
\wedge X_{i_{s+1}}\wedge\dots\wedge X_{i_n}\quad ,
\cr
&
\partial_s \wedge^n(G)=0\quad \hbox{for}\ s>n\quad ,
\cr
&
\partial_s(X_{1}\wedge\dots\wedge X_{s})=[X_{1},\dots ,X_{s}]
\quad .
\cr}
\label{codeb}
\end{equation}
\end{definition}

\begin{proposition}
\label{prop9.1}
The coderivation (\ref{codeb}) is nilpotent,
\ie, $\partial^2_s\equiv 0$.
\end{proposition}

\begin{proof}
Let $n$ and $s$ be such that $n-(s-1)\geq s$ (otherwise
the statement is trivial). Then,
\begin{equation}
  \eqalign{
     &  \partial_s\partial_s(X_1\wedge\dots\wedge X_n)
\cr
& \quad =\frac{1}{s!}\frac{1}{(n-s)!}
    \varepsilon^{i_1\dots i_n}_{1\dots n}
   \varepsilon^{j_{s+1}\dots j_n}_{i_{s+1}\dots i_n}
\big\{s
\left[ X_{j_{s+1}},\dots,X_{j_{2s-1}},\left[X_{i_1},\dots,X_{i_s}\right]\right]
\wedge X_{j_{2s}}\wedge\dots X_{j_n}
\cr
& \quad
-(n-s)[X_{j_{s+1}},\dots,X_{j_{2s}}]\wedge [X_{i_1},\dots,X_{i_s}]\wedge
X_{j_{2s+1}}\wedge\dots\wedge X_{j_n}\big\}=0\quad .\cr}
\label{codec}
\end{equation}
The first term vanishes because $s$ is even and is proportional to the
GJI.
The second one is also zero because the wedge
product of the two $s$-brackets is antisymmetric while the resulting
$\varepsilon$ symbol is
symmetric under the interchange $(i_1,\dots i_s)\leftrightarrow (j_{s+1},
\dots ,j_{2s})$, \emph{q.e.d.}
\end{proof}
\begin{remark}
A derivation satisfies Leibniz's rule (see Proposition~\ref{prop10.2} below),
which we may express as
$d\circ m = m\circ (d\otimes 1 + 1\otimes d)$
acting on the product $m$ of two copies of the algebra.
The coderivation satisfies the dual property
$\Delta\circ\partial=(\partial\otimes 1 + 1\otimes\partial)\circ \Delta$,
where $\Delta$ is the `coproduct'.
The simplest example corresponds to
\begin{equation}
\begin{array}{@{}r@{}l}
(\Delta\circ\partial)(X_1\wedge X_2) & = \Delta(\partial(X_1\wedge X_2))=
\Delta [X_1,X_2] =
[X_1,X_2]\wedge 1 + 1\wedge [X_1,X_2]=
\\[0.3cm]
& =
(\partial\otimes 1 + 1\otimes\partial)
(2 X_1\wedge 1\wedge X_2 + X_1\wedge X_2\wedge 1 + 1\wedge X_1\wedge X_2)
\end{array}
\end{equation}
since $\Delta(X_1\wedge X_2)=\Delta X_1\wedge X_2 + X_1\wedge \Delta X_2$.
\end{remark}
Let us now see how the nilpotency
condition (or equivalently the GJI) looks like in the simplest cases.

\begin{example}
\label{example9.1}
Consider $\partial\equiv\partial_2$. Then we have
\begin{equation}
     \partial(X_1\wedge X_2\wedge X_3)=
[X_1,X_2]\wedge X_3 - [X_1,X_3]\wedge X_2 + [X_2,X_3]\wedge X_1
\label{coded}
\end{equation}
and
\begin{equation}
    \partial^2(X_1\wedge X_2\wedge X_3)=
[[X_1,X_2],X_3] - [[X_1,X_3],X_2] + [[X_2,X_3],X_1] =0
\quad.                     \label{codee}
\end{equation}
\end{example}

\begin{example}
\label{example9.2}
When we move to $\partial\equiv\partial_4$, the number of terms grows very
rapidly.
The explicit expression for $\partial^2(X_{i_1}\wedge\dots\wedge X_{i_7})=0$
(which, as we
know, is equivalent to the GJI) is given in
\cite[eq. (32)]{Azc.Izq.Bue:97} (note that the tenth term there should read
$[[ X_{i_1},X_{i_2},X_{i_6},X_{i_7}],X_{i_3},X_{i_4},X_{i_5}]$).
It contains ${7\choose 3}=35$ terms. In general, the GJI which follows from
$\partial^2_{2m-2}(X_1\wedge\dots\wedge X_{4m-5})=0$ $(s=2m-2)$ contains
$4m-5\choose 2m-1$ different terms.
\end{example}

These higher order Lie algebras turn out to be a special example of the
strongly homotopy (SH) Lie algebras \cite{Lad.Sta:93,Lad.Mar:95,Jon:90}.
These allow for violations of the generalized Jacobi identity,
which are absent in our case (for the physical relevance of multialgebras, see
the references in \cite{Lad.Sta:93,Azc.Bue:97}).

\begin{definition}
[{\normalfont\textit{Strongly homotopy Lie algebras} \cite{Lad.Sta:93}}{}]
A \emph{SH Lie structure} on a vector space $V$ is a collection of
skew-symmetric linear maps
$l_n:V\otimes \mathop{\cdots}\limits^n \otimes V\to V$
such that
\begin{equation}
\sum_{i+j=n+1} \sum_{\sigma \in S_n}
{1\over (i-1)!} {1\over j!}
(-1)^{\pi(\sigma)}
(-1)^{i(j-1)} \,l_i (l_j
(v_{\sigma (1)}\otimes \dots \otimes v_{\sigma (j)}) \otimes v_{\sigma(j+1)}
\otimes \dots\otimes v_{\sigma (n)})=0 \quad.
\label{shalgebra}
\end{equation}
For a general treatment of SH Lie algebras including $v$ gradings see
\cite{Lad.Sta:93,Lad.Mar:95,Jon:90}
and references therein.
Note that
$\displaystyle {1\over (i-1)!} {1\over j!} \sum_{\sigma \in S_n}$ is equivalent
to the sum over the `unshuffles', \ie, over the permutations $\sigma\in S_n$
such that $\sigma(1)<\dots <\sigma(j)$ and $\sigma(j+1)<\dots <\sigma(n)$.
\end{definition}

\begin{example}
For $n=1$, eq. (\ref{shalgebra}) just says that $l_1^2=0$
($l_1$ is a differential).
For $n=2$, eq. (\ref{shalgebra}) gives
\begin{equation}
-{1\over 2} l_1( l_2(v_1\otimes v_2) - l_2(v_2\otimes v_1))
+l_2(l_1(v_1)\otimes v_2 - l_1(v_2)\otimes v_1)=0
\end{equation}
\ie, $l_1[v_1,v_2]=[l_1 v_1,v_2] + [v_1,l_1 v_2]$ with
$l_2(v_1\otimes v_2) = [v_1, v_2]$.

For $n=3$, we have three maps $l_1\,,l_2\,,l_3$, and eq. (\ref{shalgebra})
reduces to
\begin{equation}
\begin{array}{c}
\left[l_2 ( l_2 (v_1\otimes v_2)\otimes  v_3)
+ l_2 ( l_2 (v_2\otimes  v_3)\otimes  v_1)
+ l_2 ( l_2 (v_3\otimes  v_1)\otimes  v_2)
\right]
+\left[ l_1 ( l_3 (v_1\otimes  v_2 \otimes  v_3) ) \right]
\\[0.3cm]
+
\left[
l_3 (l_1 (v_1)\otimes  v_2 \otimes  v_3)
+ l_3 (l_1 (v_2)\otimes  v_3 \otimes  v_1)
+ l_3 (l_1 (v_3)\otimes  v_1 \otimes  v_2)
\right] = 0
\quad,
\end{array}
\end{equation}
\ie, adopting the convention that
$l_n(v_1\otimes\dots\otimes v_n)= [v_1,\dots,v_n]$,
\begin{equation}
\begin{array}{l}
[[v_1, v_2], v_3]
+ [[v_2, v_3], v_1]
+ [[v_3, v_1], v_2]
\\[0.3cm]
\qquad\qquad
=
-l_1 [v_1, v_2 , v_3]
-[l_1 (v_1), v_2 , v_3]
- [v_1, l_1 (v_2), v_3 ]
- [v_1 , v_2, l_1 (v_3)]
\quad.
\end{array}
\label{shexample}
\end{equation}
The second line in (\ref{shexample}) shows the violation of the (standard)
Jacobi identity given in the first line.
\end{example}
In the particular case in which a unique $l_n$ ($n$ even) is defined, we
recover Def.~\ref{def8.2} of a higher order Lie algebra since,
for $i=j=n$ eq. (\ref{shalgebra})
reproduces the GJI (\ref{multie}) in the form
\begin{equation}
\sum_{\sigma \in S_{2n-1}}
{1\over n!} {1\over (n-1)!} (-1)^{\pi(\sigma)}
l_n ( l_n (v_{\sigma (1)}\otimes \dots \otimes v_{\sigma (n)}) \otimes
v_{\sigma(n+1)}
\otimes \dots\otimes v_{\sigma (2n-1)})=0\quad.
\label{shalgebra1}
\end{equation}

We wish to conclude this subsection by pointing out that $n$-algebras have
also been considered in \cite{Han.Wac:95,Gne:95,Gne:96}.

\subsection{The complete BRST operator for a simple Lie algebra}
\label{sec10}

We now generalize the BRST operator and MC equations of Sec.~\ref{sec5}
to the general case of
higher-order simple Lie algebras.
The result is a new BRST-type operator that contains the
information of all the $l$ possible algebras associated with a given simple
Lie algebra $\g$ of rank $l$.

Let us first note that, in the notation of (\ref{3canonMC}), the JI reads
\begin{equation}
d^2\theta=-d(\theta\wedge\theta)=
{1\over 2}\left[ [\theta,\theta],\theta\right] =0
\quad ,
\label{completea}
\end{equation}
and expresses the nilpotency of $d$. Now, in Sec.~\ref{sec9} we considered
higher-order coderivations which also had the property
$\partial^2_s=0$
as a result of the GJI.
We may now introduce
the corresponding dual higher-order derivations ${\tilde d}_s$ to
provide a generalization of the Maurer-Cartan equations (\ref{3MC}).
Since $\partial_s$ was defined on multivectors that are product of
left-invariant
vector fields, the dual ${\tilde d}_s$ will be given for left-invariant
forms.

It is easy to introduce dual basis in $\wedge_n$ and $\wedge^n$.
With $\omega^i (X_j) =\delta_j^i$, a pair of dual basis in $\wedge_n$,
$\wedge^n$ are given by
$\omega^{I_1}\wedge\dots\wedge\omega^{I_n}$,
${1\over n!} X_{I_1}\wedge\dots\wedge X_{I_n}$ ($I_1< \dots <I_n$)
since
$(\varepsilon_{j_1\dots j_n}^{i_1\dots i_n}
\omega^{j_1}\otimes\dots\otimes\omega^{j_n})
({1\over n!} \varepsilon_{l_1\dots l_n}^{k_1\dots k_n}
X_{k_1}\otimes\dots\otimes X_{k_n})=
\varepsilon_{l_1\dots l_n}^{i_1\dots i_n}
$
and $\varepsilon_{L_1\dots L_n}^{I_1\dots I_n}$ is 1 if all indices coincide
and 0 otherwise.
Nevertheless it is customary to use the non-minimal set
$\omega^{i_1}\wedge\dots\wedge\omega^{i_n}$ to write
$\alpha={1\over n!}\alpha_{i_1\dots i_n}
\omega^{i_1}\wedge\dots\wedge\omega^{i_n}$.
Since
$(\omega^{i_1}\wedge\dots\wedge\omega^{i_n}) (X_{j_1},\dots,X_{j_n})
=\varepsilon_{j_1\dots j_n}^{i_1\dots i_n}$ it is clear that
$\alpha_{i_1\dots i_n}=\alpha(X_{i_1},\dots,X_{i_n})=
{1\over n!} \alpha(X_{i_1}\wedge\dots\wedge X_{i_n})$.

\begin{definition}
The action of ${\tilde d}_m:\wedge_n\rightarrow \wedge_{n+(2m-3)}$ (remember
that $s=2m-2$) on $\alpha\in\wedge_n$ is given by (cf. (\ref{3diffa}))
\begin{equation}
\begin{array}{r@{}l}
\displaystyle
& ({\tilde d}_m\alpha)(X_{i_1},\dots, X_{i_{n+2m-3}}):=
\\
&\qquad
\displaystyle
\frac{1}{(2m-2)!} \frac{1}{(n-1)!}
     \varepsilon_{i_1\dots i_{n+2m-3}}^{j_1\dots j_{n+2m-3}}
    \alpha([X_{j_1},\dots ,X_{j_{2m-2}}], X_{j_{2m-1}}, \dots,
      X_{j_{n+2m-3}})
\quad,
\\[0.3cm]
& \displaystyle
({\tilde d}_m\alpha)_{i_1\dots i_{n+2m-3}}
=\frac{1}{(2m-2)!} \frac{1}{(n-1)!}
\varepsilon_{i_1\dots i_{n+2m-3}}^{j_1\dots j_{n+2m-3}}
{\Omega_{j_1\dots j_{2m-2}}}^\rho_{\cdot}
\alpha_{\rho j_{2m-1}\dots j_{n+2m-3}}
\quad.
\end{array}
\label{completeb}
\end{equation}

\end{definition}

\begin{proposition}
\label{prop10.1}
$\widetilde d_m$
is dual to the coderivation $\partial_{2m-2}:\wedge^n \to \wedge^{n-(2m-3)}$,
(${\tilde d}_2=-d$, ${\tilde d}_2:\wedge_n\to \wedge_{n+1}$).
\end{proposition}

\begin{proof}
We have to check the `duality' relation  ${\tilde d}_m\alpha \propto \alpha
\partial_{2m-2}$
($\partial_{2m-2}:\wedge_{n+(2m-3)}\to \wedge_{n}$).
Indeed, if $\alpha$ is an $n$-form, eq. (\ref{codeb}) tells us that
\begin{equation}
\eqalign{
& \alpha\left( \partial_{2m-2}(X_{i_1}\wedge\dots\wedge X_{i_{n+2m-3}})\right)
=\frac{1}{(2m-2)!} \frac{1}{(n+2m-3-2m+2)!} \times
\cr
& \quad   \times \varepsilon_{i_1\dots i_{n+2m-3}}^{j_1\dots j_{n+2m-3}}
   \alpha([X_{j_1},\dots ,X_{j_{2m-2}}]\wedge X_{j_{2m-1}}\wedge \dots \wedge
      X_{j_{n+2m-3}})\quad ,}
\label{completec}
\end{equation}
which is proportional\footnote{
One finds ${\tilde d}_m\alpha= {(n+2m-3)!\over n!}\alpha\partial_{2m-2}$,
where $n$ is the order of the form $\alpha$.
The factor appears as a consequence of using the same definition
(antisymmetrization with no weight factor) for the $\wedge$ product of forms
and vectors.}
to $({\tilde d}_m\alpha)(X_{i_1}\wedge\dots\wedge
X_{i_{n+2m-3}})$, \emph{q.e.d.}
\end{proof}

\begin{proposition}
\label{prop10.2}
The operator ${\tilde d}_m$ satisfies Leibniz's rule.
\end{proposition}

\begin{proof}
For
$\alpha\in\wedge_n$, $\beta\in\wedge_p$ we get, using (\ref{completeb})
\begin{equation}
\begin{array}{r@{}l}
{\tilde d}_m (\alpha\wedge\beta) & _{i_1\dots i_{n+p+2m-3}} =
\displaystyle
{1\over (2m-2)!}{1\over (n+p-1)}
\varepsilon_{i_1\dots i_{n+p+2m-3}}^{j_1\dots j_{n+p+2m-3}}
{\Omega_{j_1\dots j_{2m-2}}}^\rho_\cdot
\\[0.3cm]
&
\displaystyle
\quad\qquad
\cdot
\Big( {1\over n!p!}
\varepsilon^{k_1\dots\dots\dots k_{n+p}}_{\rho j_{2m-1}\dots j_{n+p+2m-3}}
\alpha_{k_1\dots k_n}\beta_{k_{n+1}\dots k_{n+p}}  \Big)
\\[0.3cm]
&
\displaystyle
= {1\over (2m-2)!} {1\over n!p!}
\varepsilon_{i_1\dots i_{n+p+2m-3}}^{j_1\dots j_{n+p+2m-3}}
{\Omega_{j_1\dots j_{2m-2}}}^\rho_\cdot
\Big(
n \alpha_{\rho j_{2m-1} \dots j_{n+2m-3}}
\beta_{j_{n+2m-2} \dots j_{n+p+2m-3}}
\\[0.3cm]
&
\displaystyle
\quad\qquad
+ (-1)^n p \alpha_{j_{2m-1} \dots j_{n+2m-2}}
\beta_{\rho j_{n+2m-1}\dots j_{n+p+2m-3}} \Big)
\\[0.3cm]
&
\displaystyle
= \varepsilon_{i_1\dots i_{n+p+2m-3}}^{j_1\dots j_{n+p+2m-3}}
\Big( {1\over p! (n+2m-3)!}
({\tilde d}_m\alpha)_{j_1\dots j_{n+2m-3}} \beta_{j_{n+2m-2}\dots j_{n+p+2m-3}}
\\[0.3cm]
&
\displaystyle
\quad\qquad
+ (-1)^n {1\over n! (p+2m-3)!}
\alpha_{j_{2m-1} \dots j_{n+2m-2}}
({\tilde d}_m\beta)_{j_1\dots j_{2m-2} j_{n+2m-1}\dots j_{n+p+2m-3}} \Big)
\\[0.3cm]
&
\displaystyle
=
\Big(({\tilde d}_m\alpha)\wedge\beta + (-1)^n \alpha \wedge
({\tilde d}_m\beta)\Big)_{i_1\dots i_{n+p+2m-3}}\quad.
\end{array}
\end{equation}
Thus, ${\tilde d}_m$ is odd and
${\tilde d}_m(\alpha\wedge\beta)={\tilde d}_m\alpha
\wedge\beta+(-1)^n\alpha\wedge{\tilde d}_m\beta$, \emph{q.e.d.}
\end{proof}

The coordinates of ${\tilde d}_m\omega^\sigma$ are given by
\begin{equation}
   \eqalign{
({\tilde d}_m\omega^\sigma)(X_{i_1}, \dots, X_{i_{2m-2}})&=
\frac{1}{(2m-2)!}\varepsilon_{i_1\dots i_{2m-2}}^{j_1\dots j_{2m-2}}
\omega^\sigma([X_{j_1},\dots,X_{j_{2m-2}}])
\cr
&
=\omega^\sigma([X_{i_1},\dots,X_{i_{2m-2}}])
=\omega^\sigma({\Omega_{i_1\dots i_{2m-2}}}^\rho_\cdot X_\rho)=
{\Omega_{i_1\dots i_{2m-2}}}^\sigma_\cdot \cr}     \label{completed}
\end{equation}
from which we conclude that
\begin{equation}
      {\tilde d}_m\omega^\sigma=\frac{1}{(2m-2)!}
      {\Omega_{i_1\dots i_{2m-2}}}^\sigma_\cdot
   \omega^{i_1}\wedge\dots\wedge\omega^{i_{2m-2}}\quad .  \label{completee}
\end{equation}
For $m=2,\ \tilde d_2=-d$, equations (\ref{completee}) reproduce the MC eqs.
(\ref{3canonMC}).
In the compact notation that uses the canonical
one-form $\theta$, we may now introduce the following

\begin{proposition}
[{\normalfont\textit{Generalized Maurer-Cartan equations}}{}]
The action of ${\tilde d}_m$ on the canonical form $\theta$ is given by
\begin{equation}
{\tilde d}_m\theta=\frac{1}{(2m-2)!}
\left[ \theta,\mathop{\cdots}\limits^{2m-2},\theta \right]
\quad,
\label{completef}
\end{equation}
where the multibracket of forms is defined by
$\left[ \theta,\mathop{\cdots}\limits^{2m-2},\theta \right]=
\omega^{i_1}\wedge\dots\wedge\omega^{i_{2m-2}}[X_{i_1},\dots,X_{i_{2m-2}}]$.
Using Leibniz's rule for the operator ${\tilde d}_m$ we arrive at
\begin{equation}
{\tilde d}_m^2\theta=
-{1\over(2m-2)!}{1\over (2m-3)!}
\left[ \theta,\mathop{\cdots}\limits^{2m-3} ,\theta,
\left[ \theta,\mathop{\cdots}\limits^{2m-2} ,\theta \right]\right]
=0
\quad,
\label{completeg}
\end{equation}
which again expresses the GJI.

Each Maurer-Cartan-like
equation (\ref{completeg}) can be expressed in terms of the ghost variables
introduced in Sec.~\ref{sec5} by means of a `generalized BRST operator',
\begin{equation}
    {s}_{2m-2}=-\frac{1}{(2m-2)!}c^{i_1}\dots c^{i_{2m-2}}
     {\Omega_{i_1\dots i_{2m-2}}}^\sigma_\cdot \frac{\partial}{\partial
c^\sigma}
               \quad .                                    \label{completeh}
\end{equation}
\end{proposition}
By adding together all the $l$ generalized BRST operators, the complete
BRST operator is obtained.
Then we have the following
\begin{theorem}[{\normalfont\textit{Complete BRST operator}}{}]
\label{th10.1}
Let $\g$ be a simple Lie algebra. Then, there exists a nilpotent associated
operator, the complete BRST operator associated with \g,
given by the odd vector field
\begin{equation}
   \eqalign{
       s&=-\frac{1}{2}c^{j_1}c^{j_2}{\Omega_{j_1j_2}}^\sigma_\cdot
    \frac{\partial}{\partial c^\sigma}-\dots-
      \frac{1}{(2m_i-2)!}c^{j_1}\dots c^{j_{2m_i-2}}
     {\Omega_{j_1\dots j_{2m_i-2}}}^\sigma_\cdot \frac{\partial}{\partial
c^\sigma}
    -\dots\cr
    &-\frac{1}{(2m_l-2)!}c^{j_1}\dots c^{j_{2m_l-2}}
     {\Omega_{j_1\dots j_{2m_l-2}}}^\sigma_\cdot \frac{\partial}{\partial
c^\sigma}
    \equiv s_2+\dots +s_{2m_i-2}+\dots +s_{2m_l-2}\quad , \cr}
                                                          \label{completei}
\end{equation}
where $i=1,\dots,l$, ${\Omega_{j_1j_2}}^\sigma_\cdot\equiv
{C_{j_1j_2}}^\sigma_\cdot$ and ${\Omega_{j_1\dots j_{2m_i-2}}}^\sigma_\cdot$
are the corresponding $l$ higher-order cocycles.
\end{theorem}

\begin{proof}
We have to show that $\{ s_{2m_i-2},s_{2m_j-2}\}=0$
$\forall\, i,j$. To prove it, let us write the anti-commutator explicitly:
\begin{equation}
\eqalign{
&
\{ s_{2m_i-2},s_{2m_j-2}\}=
\frac{1}{(2m_i-2)!}\frac{1}{(2m_j-2)!}
\times
\cr
&\quad
\times
\{ (2m_j-2)c^{l_1}\dots c^{l_{2m_i-2}}
{\Omega_{l_1\dots l_{2m_i-2}}}^\rho_\cdot
c^{r_2}\dots c^{r_{2m_j-2}}
{\Omega_{\rho r_2\dots r_{2m_j-2}}}^\sigma_\cdot\frac{\partial}{\partial
c^\sigma}
+i\leftrightarrow j
\cr
&\quad
+(c^{l_1}\dots c^{l_{2m_i-2}}c^{r_1}\dots c^{r_{2m_j-2}}
{\Omega_{l_1\dots l_{2m_i-2}}}^\rho_\cdot
{\Omega_{ r_1\dots r_{2m_j-2}}}^\sigma_\cdot+i\leftrightarrow j)
\frac{\partial}{\partial c^\rho}\frac{\partial}{\partial c^\sigma}\}
\cr
&\quad
=\frac{1}{(2m_i-2)!}\frac{1}{(2m_j-3)!}c^{l_1}\dots c^{l_{2m_i-2}}
c^{r_2}\dots c^{r_{2m_j-2}}{\Omega_{l_1\dots l_{2m_i-2}}}^\rho_\cdot
{\Omega_{\rho r_2\dots r_{2m_j-2}}}^\sigma_\cdot\frac{\partial}{\partial
c^\sigma}
+i\leftrightarrow j\ ,
\cr}
\label{completej}
\end{equation}
where we have used the fact that $\frac{\partial}{\partial
c^\rho}\frac{\partial}{\partial c^\sigma}$
is antisymmetric in $\rho,\sigma$ while the parenthesis multiplying it is
symmetric. The term proportional to a single $\frac{\partial}{\partial
c^\sigma}$ also
vanishes as a consequence of equation (\ref{multij}), \emph{q.e.d.}
\end{proof}

The coefficients of $\partial / \partial c^\sigma$
in $s_{2m_i-2}$ can be viewed, in dual terms, as (even)
multivectors of the type
\begin{equation}
\Lambda=\frac{1}{(2m-2)!}{\Omega_{i_1\dots i_{2m-2}}}^\sigma_\cdot x_\sigma
\partial^{i_1}\wedge\dots \wedge \partial^{i_{2m-2}}\quad .
\label{completek}
\end{equation}
(see (\ref{lambda})).
They have the property of having zero Schouten-Nijenhuis bracket among
themselves by virtue of the GJI (\ref{multih}).

\begin{definition}
\label{def10.2}
Let us consider the algebra $\wedge(M)$ of multivectors on $M$. The
Schouten-Nijenhuis bracket (SNB) of $A\in \wedge^p(M)$ and $B\in \wedge^q(M)$
is the unique extension of the Lie bracket of two vector fields to a bilinear
mapping $\wedge^p(M)\times\wedge^q(M)\rightarrow \wedge^{p+q-1}(M)$ in
such a way that $\wedge(M)$ becomes a graded superalgebra.
\end{definition}

For the expression of the SNB in coordinates we refer to \cite{Nij:55,Lic:77}.
It turns out that the multivector algebra with the exterior product and the SNB
is a Gerstenhaber algebra\footnote{
A Gerstenhaber algebra \cite{Ger:63} is a ${\mathbb Z}$-graded vector space
(with homogeneous subspaces $\wedge^a$, $a$ being the grade) with two bilinear
multiplication operators, $\cdot$ and $[\ ,\ ]$ with the following
properties ($u\in \wedge^a$, $v\in \wedge^b$, $w\in \wedge^c$):
\begin{quote}
a) deg$(u\cdot v)=a+b$,
\\
b) deg$[u,v]=a+b-1$,
\\
c) $(u\cdot v)\cdot w=u\cdot (v\cdot w)$,
\\
d) $[u,v]=-(-1)^{(a-1)(b-1)}[v,u]$,
\\
e) $(-1)^{(a-1)(c-1)}[u,[v,w]]+(-1)^{(c-1)(b-1)}[w,[u,v]]+
(-1)^{(b-1)(a-1)}[v,[w,u]]=0$,
\\
f) $[u,v\cdot w]=[u,v]\cdot w+(-1)^{(a-1)b}v\cdot [u,w]$.
\end{quote}
For an analysis of various related algebras, including Poisson algebras, see
\cite{Kos:96} and references therein.},
in which deg$(A)=p-1$ if $A\in \wedge^p$. Thus, the multivectors of the
form (\ref{completek}) form an abelian subalgebra of this Gerstenhaber
algebra, the commutativity (in the sense of the SNB)
being a consequence of (\ref{multih}).

\section{Higher order generalized Poisson structures}
\label{sec11}

We shall consider in this section two possible generalizations of the ordinary
Poisson structures (PS) by brackets of more than two functions.
The first one is the Nambu-Poisson structure (N-P)
\cite{Nam:73,Sah.Val:92,Tak:94,Fil:85} (see also \cite{Mic.Vin:98}).
The second, named generalized PS (GPS) \cite{Azc.Per.Bue:96b,Azc.Per.Bue:96},
is based on the previous constructions (and has been extended to the
supersymmetric case \cite{Azc.Izq.Per.Bue:97}).
We shall present both generalizations as
well as examples of the GPS, which are naturally obtained from
the higher-order simple Lie algebras of Sec.~\ref{sec8}.
A comparison between both structures may be found in \cite{Azc.Izq.Bue:97}
and in table~\ref{table11.1} (see also
\cite{Iba.Leo.Mar.Die:97}).
Let us first review briefly the standard PS.

\subsection{Standard Poisson structures}
\label{sec11.1}

\begin{definition}
\label{def11.1}
Let $M$ be a differentiable manifold. A Poisson bracket (PB) on ${\cal F}(M)$
is a bilinear mapping
$\{\cdot,\cdot\} :{\cal F}(M)\times {\cal F}(M)\rightarrow
{\cal F}(M)$ that satisfies ($f,g,h\in {\cal F}(M)$)
\begin{description}
\item[a)] Skew-symmetry
\begin{equation}
\{ f,g\}=-\{ g,f\}\quad,
\end{equation}
\item[b)] Leibniz's rule,
\begin{equation}
\{ f,gh\}=g\{ f,h\}+\{ f,g\}h\quad,
\end{equation}
\item[c)] Jacobi identity
\begin{equation}
\hbox{Alt}
\{ f,\{ g, h\}\}= \{ f,\{ g, h\}\} + \{ g,\{ h, f\}\} + \{ h,\{ f, g\}\} =0
 \quad.
\label{PSji}
\end{equation}
\end{description}
A PB on $M$ defines a PS.
\end{definition}

In local coordinates $\{ x^i\}$, conditions a), b) and c) mean that it is
possible to write
\begin{equation}
    \{ f(x),g(x)\}=\omega^{ij}\partial_i f\partial_j g
\quad,\quad
\omega^{ij}=-\omega^{ji}
\quad,\quad
   \omega^{jk}\partial_k \omega^{lm}+\omega^{lk}\partial_k \omega^{mj}
   +\omega^{mk}\partial_k \omega^{jl}=0\quad .             \label{GPSb}
\end{equation}
It is possible to rewrite a)-c) in a geometrical way by using the bivector
\begin{equation}
    \Lambda=\frac{1}{2}\omega^{jk}\partial_j\wedge\partial_k \quad ,
\label{GPSc}
\end{equation}
in terms of which
\begin{equation}
   \{ f,g\}=\Lambda(df,dg) \quad;
\label{GPSd}
\end{equation}
the JI imposes a condition on $\Lambda$, which is equivalent to the vanishing
of the SNB \cite{Lic:77}
\begin{equation}
[\Lambda,\Lambda]=0 \quad   .                         \label{GPSe}
\end{equation}
If the manifold $M$ is the dual of a Lie algebra, there always exists a PS, the
Lie-Poisson structure, which is obtained by defining the fundamental
Poisson bracket $\{ x_i,x_j \}$ (where $\{ x_i\}$ are coordinates on $\g^*$).
Since $\g\sim (\g^*)^*$, we may think of $\g$ as a
subspace
of the ring of smooth functions ${\cal F}(\g^*)$.
Then, the Lie algebra
commutation relations
\begin{equation}
    \{ x_i,x_j\}=C^k_{ij}x_k                         \label{GPSfb}
\end{equation}
define, by assuming b) above, a mapping ${\cal F}(\g^*)\times {\cal F}(\g^*)
\rightarrow {\cal F}(\g^*)$ associated with the bivector
$\Lambda=\frac{1}{2}C^k_{ij}x_k
\frac{\partial}{\partial x_i}\wedge\frac{\partial}{\partial x_i}$.
This is a PB since condition (\ref{GPSb}) (or (\ref{GPSe})) is equivalent to
the JI for the structure constants of \g.

\subsection{Nambu-Poisson structures}
\label{sec11.2}

Already in 1973, Nambu \cite{Nam:73} considered the possibility of extending
Poisson brackets to brackets of three functions. His attempt has been
generalized since then, and all generalizations considered share the
following two properties
\begin{equation}
\begin{array}{l}
\mbox{a)}\ \{ f_1,\dots,f_i,\dots,f_j,\dots, f_n\}=-\{ f_1,\dots,f_j,\dots,f_i,
\dots, f_n\}\ \mbox{(skew-symmetry)}\ ,
\\
\mbox{b)}\ \{f_1,\dots,f_{n-1},gh\}=g\{f_1,\dots,f_{n-1},h\}+
\{f_1,\dots,f_{n-1},g\}h\ \mbox{(Leibniz's rule)}
\end{array}
\label{GPSfc}
\end{equation}
which will be guaranteed if the bracket is generated in local coordinates
$\{ x_i\}$ on $M$ by
\begin{equation}
     \Lambda=\frac{1}{n!}\eta_{i_1\dots i_n}\partial^{i_1}\wedge\dots\wedge
   \partial^{i_n}                                                \label{GPSf}
\end{equation}
as in (\ref{GPSd}), \ie\ by
\begin{equation}
     \{f_1,\dots,f_n\}=\Lambda(df_1,\dots,df_n)\quad .        \label{GPSg}
\end{equation}

The key difference among the higher order PS is the
identity that generalizes c) in Definition~\ref{def11.1}.
That corresponding to Nambu's mechanics was given by Sahoo and Valsakumar
\cite{Sah.Val:92} and in the
general case by Takhtajan \cite{Tak:94}, who studied it in detail and named it
the \emph{fundamental identity} (FI)
\begin{equation}
  \eqalign{
&   \{ f_1,\dots,f_{n-1},\{ g_1,\dots,g_n\}\}=\{\{ f_1,\dots,f_{n-1},g_1\},
     g_2,\dots,g_n\}\cr
&\qquad +\{g_1,\{ f_1,\dots,f_{n-1},g_2\},g_3,\dots,g_n\}+\dots+
   \{ g_1\dots,g_{n-1},\{ f_1,\dots,f_{n-1},g_n\}\}\cr}   \label{GPSh}
\end{equation}
(see also \cite{Fil:85,Mic.Vin:98}).
The FI (\ref{GPSh}), together with (\ref{GPSfc}), define the Nambu-Poisson
structures \cite{Tak:94}.
To see the signification of (\ref{GPSh}), let us consider
$n-1$ `Hamiltonians' $(H_1,\dots, H_{n-1})$ and define the time evolution of an
observable by
\begin{equation}
    {\dot g}=\{ H_1,\dots, H_{n-1},g\}\quad .              \label{GPSi}
\end{equation}
Then, the FI guarantees that
\begin{equation}
   \frac{d}{dt}\{ g_1,\dots,g_n\}=\{ {\dot g}_1,\dots,g_n\}+\dots +
  \{ g_1,\dots,{\dot g}_n\}\quad ,                           \label{GPSj}
\end{equation}
\ie, that the time derivative is a derivation of the
N-P $n$-bracket.
In this way, the bracket of any $n$ constants of the motion is itself a
constant of the motion.

Inserting (\ref{GPSf}) into (\ref{GPSh}), one gets two conditions
\cite{Tak:94} for the coordinates $\eta_{i_1\dots i_n}$ of $\Lambda$.
The first is the \emph{differential condition},
which in local coordinates may be written as
\begin{equation}
\eta_{i_1\dots i_{n-1}\rho}\partial^\rho\eta_{j_1\dots j_n}
-{1\over (n-1)!}\epsilon^{l_1\dots l_n}_{j_1\dots j_n}
(\partial^\rho\eta_{i_1\dots i_{n-1} l_1})\eta_{\rho l_2\dots l_n}
=0 \quad.
\label{difcond}
\end{equation}
The second is the \emph{algebraic condition}.
It follows from requiring the
vanishing of the second derivatives in (\ref{GPSh}). In local coordinates
it reads
\begin{equation}
\Sigma + P(\Sigma)=0\quad,
\label{alcond}
\end{equation}
where $\Sigma$ is the $2n$-tensor
\begin{equation}
\begin{array}{rl}
\Sigma_{i_1\dots i_n j_1\dots j_n}
=&\eta_{i_1\dots i_{n}}\eta_{j_1\dots j_{n}}
-\eta_{i_1\dots i_{n-1}j_1}\eta_{i_n j_2 \dots j_{n}}
-\eta_{i_1\dots i_{n-1}j_2}\eta_{j_1 i_n j_3\dots j_{n}}
\\[0.2cm]
-& \eta_{i_1\dots i_{n-1}j_3}\eta_{j_1 j_2 i_n j_4 \dots j_{n}}-
\dots
-\eta_{i_1\dots i_{n-1}j_n}\eta_{ j_1j_2 \dots j_{n-1} i_n} \;.
\end{array}
\label{Sigmadef}
\end{equation}
It turns out \cite{Ale.Guh:96,Gau:96,Hie:97}
(see also \cite{Dit.Fla.Ste.Tak:97})
that this last condition implies that $\Lambda$ in
(\ref{GPSf}) is decomposable,
\ie, that $\Lambda$ can be written as the exterior product of vector
fields.

\subsection{Generalized Poisson structures}
\label{sec11.3}

Instead of generalizing Jacobi's identity through the FI (\ref{GPSh}),
one may take a different path by following a geometrical rather than a
dynamical approach.
Since for the ordinary PS the JI is given by (\ref{GPSe}),
it is natural \cite{Azc.Per.Bue:96b,Azc.Per.Bue:96} to introduce in the even
case new GPS by means of
\begin{definition}
A $2p$-multivector $\Lambda^{(2p)}$ defines a GPS if it satisfies
\begin{equation}
     [\Lambda^{(2p)},\Lambda^{(2p)}]=0 \quad ,                \label{GPSk}
\end{equation}
where $[\ ,\ ]$ denotes again the SNB.
Notice that for $n$ odd,
$[\Lambda^{(n)},\Lambda^{(n)}]$ vanishes identically and hence the condition is
empty.
Written in terms of the coordinates of
$\Lambda^{(2p)}$ (now denoted $\omega_{i_1\dots i_{2p}}$), the GJI
condition (\ref{GPSk}) reads
\begin{equation}
    \varepsilon^{j_1\dots j_{4p-1}}_{i_1\dots i_{4p-1}}
      \omega_{j_1\dots j_{2p-1}\sigma}\partial^\sigma \omega_{j_{2p}\dots
    j_{4p-1}}=0
\label{GPSl}
\end{equation}
\end{definition}
[cf. (\ref{multih})].
Thus a GPS is defined by (\ref{GPSfc}) and eq. (\ref{GPSk}) (or (\ref{GPSl})),
which in terms of the GPB is expressed by
\begin{proposition}
\label{prop11.1}
The GJI for the GPB
\begin{equation}
   \eqalign{
   \hbox{Alt}\{ f_1,\dots ,&f_{2p-1},\{ f_{2p},\dots,f_{4p-1}\} \}\cr
    &:=\sum_{\sigma\in S_{4p-1}}(-1)^{\pi(\sigma)}\{ f_{\sigma(1)},\dots,
    f_{\sigma(2p-1)},\{ f_{\sigma(2p)},\dots,f_{\sigma(4p-1)}\} \}
     =0 \quad,\cr}                                               \label{GPSm}
\end{equation}
is equivalent \cite{Azc.Per.Bue:96b,Azc.Per.Bue:96} to condition (\ref{GPSl}).
\end{proposition}

\begin{proof}
Let us write (\ref{GPSm}) as
\begin{equation}
\eqalign{
\varepsilon^{j_1\dots j_{4p-1}}_{i_1\dots i_{4p-1}}
&
\{ f_{j_1},\dots,f_{j_{2p-1}},\omega_{l_{2p}\dots l_{4p-1}}
\partial^{l_{2p}} f_{j_{2p}}\dots \partial^{l_{4p-1}}f_{j_{4p-1}}\}
\cr &
=
\varepsilon^{j_1\dots j_{4p-1}}_{i_1\dots i_{4p-1}}
\omega_{l_1\dots l_{2p-1}\sigma}
\partial^{l_1}f_{j_1}\dots\partial^{l_{2p-1}}f_{j_{2p-1}}
(\partial^\sigma \omega_{l_{2p}\dots l_{4p-1}}\partial^{l_{2p}}f_{j_{2p}}\dots
\partial^{l_{4p-1}}f_{j_{4p-1}}
\cr &
+ 2p\omega_{l_{2p}\dots l_{4p-1}}\partial^\sigma\partial^{2p}f_{j_{2p}}
\partial^{l_{2p+1}}f_{j_{2p+1}}\dots\partial^{l_{4p-1}}f_{j_{4p-1}})=0\quad .
\cr }
\label{GPSn}
\end{equation}
The second term vanishes because the factor multiplying
$\partial^\sigma\partial^{2p}f_{j_{2p}}$ is antisymmetric with respect to the
interchange
$\sigma\leftrightarrow l_{2p}$. Hence, we are left with (\ref{GPSl})
because $f_{j_1},\dots ,f_{j_{4p-1}}$ are arbitrary, \emph{q.e.d.}
\end{proof}

\begin{remarks}
1.~~It is also possible to define the GPS in the \emph{odd} case
\cite{Azc.Izq.Bue:97}.
For GPB with an odd number of arguments, the second
term in (\ref{GPSn}) does not vanish, giving now rise (as for the N-P
structures) to an `algebraic condition' which is absent in the even case
\cite{Azc.Izq.Bue:97}.
\\
2.~~These constructions may also be extended to the $\mathbb{Z}_2$-graded
(`supersymmetric') case \cite{Azc.Izq.Per.Bue:97}.
\\
3.~~The GJI does not imply the FI. Thus, the GPB
of constants of the motion is not a constant of the motion in general (see
\cite{Azc.Per.Bue:96}, however, for a weaker result).
On the other hand, the FI
does imply the GJI when $n$ is even (and
also when $n$ is odd). So the GPS's may be viewed as a generalization of the
Nambu-Takhtajan one.
As a result, a $\Lambda$ defining a GPS is not decomposable in general.
\end{remarks}

\subsection{Higher order linear Poisson structures}
\label{sec11.4}

It is now easy to construct examples of GPS (infinitely many, in fact) in the
linear case.
They are obtained by extending the
argument at the end of Sec.~\ref{sec11.1} to the GPS.
Let \g\ be a simple Lie algebra of rank $l$. We know from Sec.~\ref{sec8}
that
corresponding to it there are $(l-1)$ higher order Lie algebras.
Their structure constants define a GPB
$\{\cdot,\mathop{\cdots}\limits^{2m_l-2},\cdot\}:
\g^*\times \mathop{\cdots}\limits^{2m_l-2} \times\g^*\rightarrow \g^*$
by
\begin{equation}
       \{ x_{i_1},\dots, x_{i_{2m_l-2}}\}={\Omega_{i_1\dots
      i_{2m_l-2}}}^\sigma_\cdot x_\sigma\quad ,               \label{GPSo}
\end{equation}
where $\Omega$ is the $(2m_l-1)$-cocycle. If one now computes the
GJI (\ref{GPSl}) for $\omega_{i_1\dots i_{2m_l-2}}=
{\Omega_{i_1\dots i_{2m_l-2}}}^\sigma_\cdot x_\sigma$, or, alternatively,
$[\Lambda,\Lambda]$ for
\begin{equation}
\Lambda=\frac{1}{(2m-2)!}{\Omega_{i_1\dots i_{2m-2}}}^\sigma_\cdot x_\sigma
\partial^{i_1}\wedge\dots \wedge \partial^{i_{2m-2}}\quad,
\label{lambda}
\end{equation}
one sees that $[\Lambda,\Lambda]=0$ is satisfied since it expresses
the GJI for the higher order structure constants
$\Omega$ given in (\ref{multih}). This means that \emph{all}
higher-order simple Lie algebras define linear GPS.
These structures are not of the Nambu-Poisson type.

Conversely, given a linear GPS with fundamental GPB (\ref{GPSo}),
the associated higher-order Lie algebra provides a
realization of it. This is what one might
expect to achieve when quantizing the classical theory if, that is,
quantization implies
the replacement of observables by \emph{associative} operators and the GPB by
multicommutators (the standard quantization \emph{\`a la} Dirac implies the
well known substitution $\{\ ,\ \}\mapsto {1\over i\hbar} [\ ,\ ]$).
The physical difficulty for the GPS is the fact that time derivative is
not a derivation of the bracket (Sec.~\ref{sec11.3}). The
N-P structures are free from this problem, but the FI
is not an identity for the algebra of associative operators.
Thus, one is led to the conclusion that a standard quantization of higher order
mechanics
is not possible (see, however, \cite{Dit.Fla.Ste.Tak:97}) and that ordinary
Hamiltonian mechanics is, in this sense, rather unique.

\begin{table}
\begin{center}
\begin{tabular}{lccc}
& PS & N-P & GPS (even order)
\\
\hline
\hline
Characteristic identity (CI): & Eq. (\ref{PSji}) (JI) & Eq. (\ref{GPSh}) (FI) &
Eq. (\ref{GPSm}) (GJI)
\\
Defining conditions: & Eq. (\ref{GPSb}) & Eqs. (\ref{difcond}),(\ref{alcond}) &
Eq. (\ref{GPSl})
\\
Liouville theorem: & Yes & Yes & Yes
\\
Poisson theorem: & Yes & Yes & No (in general)
\\
\begin{minipage}[b]{5.0cm}
CI realization in terms of associative
operators:
\end{minipage}
& Yes & No (in general)& Yes
\end{tabular}
\caption{Some properties of Nambu-Poisson (N-P) and generalized
Poisson (GP) structures.}
\label{table11.1}
\end{center}
\end{table}

\section{Relative cohomology, coset spaces and effective WZW actions}
\label{sec13}

This is a topic of recent physical interest
\cite{DHo.Wei:94,DHo:95,Azc.Mac.Per:98} since, for an action invariant under
the compact symmetry group $G$ which has a vacuum that
is symmetric under the subgroup $H$, the Goldstone fields parametrize the
coset space.
Thus, the possible invariant effective actions of WZW type
\cite{Wit:84,Wit:83} are related with the cohomology in $G/H$.
In particular, for the cohomology of degree 4 and 5 we may construct WZW
actions on 3- and 4- dimensional space-times respectively.

Let $G$ be a compact Lie group and $H$ a subgroup. The `left
coset' $K=G/H$ is defined through the projection map $\pi:G\to K$ by
\begin{equation}
\pi: g h \to \{gH\} \equiv g \quad, \quad \forall\, h\in H
\quad.
\label{coset1}
\end{equation}
$G(H,K)$ is a principal bundle where the structure group $H$ acts on the right
$R_h: g \mapsto gh$ and the base space is the coset $G/H$.

\begin{theorem}[{\normalfont\textit{Projectable forms}}{}]
\label{th13.1}
Let $G(H,K)$ be a principal bundle.
A $q$-form $\Omega$ on $G$ is projectable to a form $\bar\Omega$ on $K$,
\ie, there exists a unique $\bar\Omega$ such that
$\Omega = \pi^* (\bar \Omega)$
iff
\begin{quote}
$\Omega(g)(X_1(g),\dots, X_q(g)) = 0$  if one $X\in {\mathfrak X}(H)$
($\Omega$ is horizontal)
\\[0.25cm]
$R_h^*\Omega = \Omega$ ($\Omega$ is invariant under the right action of
$H$).
\end{quote}
\end{theorem}

\begin{proof}
See \cite{Kob.Nom:69}.
\end{proof}

\begin{definition}[{\normalfont\textit{Relative Lie algebra cohomology}}{}]
\label{def13.1}
Let $\g$ be a Lie algebra and ${\cal H}$ a subalgebra of \g.
The space of relative (to the subalgebra ${\cal H}$) $q$-cochains
$C^q(\g,{\cal H})$ is that of the
$q$-skew-symmetric maps $\Omega:\g\wedge\mathop{\cdots}\limits^q\wedge\g\to\R$
such that (cf. Theorem~\ref{th13.1})
\begin{equation}
\begin{array}{l}
\Omega(X,X_2,\dots,X_q)=0 \quad \hbox{if}\ X\in {\cal H}\ (\Omega\
\mbox{is\ \emph{horizontal}})
\\[0.3cm]
\Omega([X,X_1],X_2,\dots,X_q) + \dots +
\Omega(X_1,X_2,\dots, [X, X_q]) = 0 \quad \forall\, X\in {\cal H} \quad.
\end{array}
\label{coset3}
\end{equation}
The cocycles and coboundaries are then defined by
\begin{equation}
Z^q(\g,{\cal H}) = Z^q(\g) \cap C^q(\g,{\cal H})
\quad,\quad
B^q(\g,{\cal H}) = s C^{q-1}(\g,{\cal H})
\label{coset5}
\end{equation}
where $s$ is the standard Lie algebra cohomology operator.
The \emph{relative Lie algebra cohomology groups} are now defined as usual,
\begin{equation}
H^q(\g,{\cal H}) = Z^q(\g,{\cal H}) / B^q(\g,{\cal H})\quad.
\label{coset6}
\end{equation}
\end{definition}

Let us consider a horizontal LI form $\Omega$ on $G$
and which is invariant under the right action of ${\cal H}$, namely
\begin{equation}
i_{X(g)} \Omega(g) = 0
\quad,\quad
L_{X(g)} \Omega(g) = 0
\quad \forall\, X\in {\cal H}
\label{coset7}
\end{equation}
Since there is a one-to-one correspondence between LI forms on $\Omega$ and
multilinear mappings on $\g$, it is clear that (\ref{coset7}) is the
translation of (\ref{coset3}) (Theorem~\ref{th13.1}) in terms of differential
forms on the group manifold $G$.

\begin{theorem}
\label{th13.2}
The ring of invariant forms on $G/H$ is given by the exterior algebra
of multilinear antisymmetric maps on $\g$ vanishing on ${\cal H}$ and
which are $ad\,{\cal H}$-invariant.
\end{theorem}

\begin{remark}
Definition~\ref{def13.1} requires to prove that $sC^{q} \subset C^{q+1}$.
But this may be seen using that (\ref{coset3}) may be
written as $i_X \Omega (X_2, \dots, X_q) =0$ and
$L_X \Omega (X_1,\dots, X_q)=0,\ X\in {\cal H}$.
Now,
\begin{equation}
i_X(s\Omega) (X_1,\dots, X_q) = (L_X -s i_X) \Omega (X_1,\dots, X_q) =0
\end{equation}
and
\begin{equation}
L_X(s\Omega) (X_1,\dots, X_q) = (s L_X) \Omega (X_1,\dots, X_q) =0
\end{equation}
since $s\sim d$, $s i_X + i_X s = L_X$ and $[L_X, s]=0$.
\end{remark}

\begin{theorem}
\label{th13.3}
The Lie algebra cohomology groups $H^q(\g, {\cal H})$ relative to ${\cal H}$
are given by the
forms $\Omega$ on $G$ which are a) LI b) closed and c) projectable.
\end{theorem}

\begin{proof}
LI means that they can be put in one-to-one correspondence with skew-linear
forms on $\g$;
closed implies that $d\Omega=0$ or, in terms of the
cohomology operator, that $s\Omega=0$.
Finally, projectable means that the relative cohomology conditions
(\ref{coset3}) are satisfied, \emph{q.e.d.}
\end{proof}

Note that, again, the relative and the de Rham cohomology on the coset may be
different.
However, if $G$ is compact the following theorem
\cite[Theorem 22.1]{Che.Eil:48}
holds
\begin{theorem}
\label{th13.4}
Let $G$ be a compact and connected Lie group, $H$ a closed connected subgroup
and $K$ the homogeneous space $K=G/H$.
Then $H^q(\g, {\cal H})$ and $H^q_{DR}(K)$ are isomorphic, and so are their
corresponding rings
$H^*(\g, {\cal H})$ and $H^*_{DR}(K)$.
\end{theorem}

The relative cohomology may be used to construct effective actions of WZW type
on coset spaces \cite{DHo.Wei:94,DHo:95,Azc.Mac.Per:98};
the obstruction may be expressed in terms of an anomaly.
For instance, when it is absent, the five cocycle on $G/H$ has the form
\begin{equation}
\mbox{Tr}({\cal U}^5) - 5 \mbox{Tr}({\cal W}{\cal U}^3) + 10
\mbox{Tr}({\cal W}^2{\cal U})
\quad,
\end{equation}
where ${\cal U}$ is the $({\cal G}\backslash{\cal H})$-component of the
canonical form
$\theta$ on $G$ and ${\cal W}=d{\cal V} + {\cal V}\wedge{\cal V}$ is the
curvature of the ${\cal H}$-valued
connection ${\cal V}$ given by the ${\cal H}$-component $\omega^\alpha$ of
$\theta$.
In fact, a similar procedure is also valid to recover the obstructions to the
process of gauging WZW actions found in \cite{Hul.Spe:91}.
It may be seen that this is due to the relation between the relative Lie
algebra cohomology and the equivariant (see \cite{Ber.Get.Ver:91})
cohomology, but we shall not develop this point here (see \cite{Azc.Per:98}
and references therein).

\subsection*{Acknowledgements}
This paper has been partially supported by a research grant from
the MEC, Spain (PB96-0756).
J.~C.~P.~B. wishes to thank the Spanish MEC and the CSIC for an FPI grant.
The authors also whish to thank J.~Stasheff for helpful correspondence.

\begin{small}

\end{small}

\end{document}